\pdfoutput=1

\documentclass[conference]{llncs}

\usepackage{subfig}
\usepackage{graphicx}
\usepackage{color}
\usepackage{hyperref}
\usepackage{multirow}
\usepackage{amsmath, amstext, amsfonts, amssymb}

\newcommand{\eat}[1]{}
\usepackage{algorithm}
\usepackage{algpseudocode}

\begin{document}

\title{I Know Why You Went to the Clinic:\\Risks and Realization of HTTPS Traffic Analysis}
\author{Brad Miller\inst{1} \and Ling Huang\inst{2} \and A. D. Joseph\inst{1} \and J. D. Tygar\inst{1}}
\institute{UC Berkeley \and Intel Labs}

\maketitle

\begin{abstract}
Revelations of large scale electronic surveillance and data mining by governments and corporations have fueled increased adoption of HTTPS.
We present a traffic analysis attack against over 6000 webpages spanning the HTTPS deployments of 10 widely used, industry-leading websites in areas such as healthcare, finance, legal services and streaming video.
Our attack identifies individual pages in the same website with 89\% accuracy, exposing personal details including medical conditions, financial and legal affairs and sexual orientation.
We examine evaluation methodology and reveal accuracy variations as large as 18\% caused by assumptions affecting caching and cookies.
We present a novel defense reducing attack accuracy to 27\% with a 9\% traffic increase, and demonstrate significantly increased effectiveness of prior defenses in our evaluation context, inclusive of enabled caching, user-specific cookies and pages within the same website.
\end{abstract}

\section{Introduction}
\label{sec:intro}

HTTPS is far more vulnerable to traffic analysis than has been previously discussed by researchers.
In a series of important papers, a variety of researchers have shown a number of traffic analysis attacks on SSL proxies~\cite{hintz,sun}, SSH tunnels~\cite{herrmann,cai,dyer,liberatore,bissias}, Tor~\cite{herrmann,cai,panchenko,wang}, and in unpublished work, HTTPS~\cite{cheng,danezis}.
Together, these results suggest that HTTPS may be vulnerable to traffic analysis.
This paper confirms the vulnerability of HTTPS, but more importantly, gives new and much sharper attacks on HTTPS, presenting algorithms that decrease errors 3.6x from the best previous techniques.
We show the following novel results:

\begin{itemize}
\item Novel attack technique capable of achieving 89\% accuracy over 500 pages hosted at the same website, as compared to 60\% with previous techniques
\item Impact of caching and cookies on traffic characteristics and attack performance, affecting accuracy as much as 18\%
\item Novel defense reducing accuracy to 27\% with 9\% traffic increase; significantly increased effectiveness of packet level defenses in the HTTPS context
\end{itemize}
\noindent We evaluate attack, defense and measurement techniques on websites for healthcare (Mayo Clinic, Planned Parenthood, Kaiser Permanente), finance (Wells Fargo, Bank of America, Vanguard), legal services (ACLU, Legal Zoom) and streaming video (Netflix, YouTube).

We design our attack to distinguish minor variations in HTTPS traffic from significant variations which indicate distinct traffic contents.
Minor traffic variations may be caused by caching, dynamically generated content, or user-specific content including cookies.
Our attack applies clustering techniques to identify patterns in traffic.
We then use a Gaussian distribution to determine similarity to each cluster and map traffic samples into a fixed width representation compatible with a wide range of machine learning techniques.
Due to similarity with the Bag-of-Words approach to document classification, we refer to our technique as Bag-of-Gaussians (BoG).
This approach allows us to identify specific pages within a website, even when the pages have similar structures and shared resources.
After initial classification, we apply a hidden Markov model (HMM) to leverage the link
structure of the website and further increase accuracy.
We show our approach achieves substantially greater accuracy than attacks developed by Panchenko \textit{et al.} (Pan)~\cite{panchenko}, Liberatore and Levine (LL)~\cite{liberatore}, and Wang \textit{et al.}~\cite{wang}.

We also present a novel defense technique and evaluate several previously proposed defenses.
We consider deployability both in the design of our technique and the selection of previous techniques.
Whereas the previous, and less effective, techniques could be implemented as stateless packet filters, our technique operates statelessly at the granularity of individual HTTP requests and responses.
Our evaluation demonstrates that some techniques which are ineffective in other traffic analysis contexts have significantly increased impact in the HTTPS context.
For example, although Dyer \textit{et al.} report exponential padding as only decreasing accuracy of the Panchenko classifier from 97.2\% to 96.6\%, we observe a decrease from 60\% to 22\%~\cite{dyer}.
Our novel defense reduces the accuracy of the BoG attack from 89\% to 27\% while generating only 9\% traffic overhead.

We conduct our evaluations using a dataset of 463,125 page loads collected from 10 websites during December 2013 and January 2014.  
Our collection infrastructure includes virtual machines (VMs) which operate in four separate collection modes, varying properties such as caching and cookie retention across the collection modes.
By training a model using data from a specific collection mode and evaluating the model using a different collection mode, we are able to isolate the impact of factors such as caching and user-specific cookies on analysis results.
We present these results along with insights into the fundamental properties of the traffic itself.

Section~\ref{sec:motivation} presents the risks posed by HTTPS traffic analysis and adversaries who may be motivated and capable to conduct attacks.
Section~\ref{sec:rel_work} reviews prior work, and in section~\ref{sec:attack_presentation} we present the core components of our attack.
Section~\ref{sec:eval} presents the impact of evaluation conditions on reported attack accuracy, section~\ref{sec:attack_eval} evaluates our attack, and section~\ref{sec:defense_eval} presents and evaluates defense techniques.
In Section~\ref{sec:conclusion} we discuss results and conclude.

\section{Risks of HTTPS Traffic Analysis}
\label{sec:motivation}
This section presents an overview of the potential risks and attackers we consider in analyzing HTTPS traffic analysis attacks.
Section~\ref{sec:https_scenarios} describes four categories of content, each of which we explore in this work, and potential consequences of a privacy violation in each category.
Section~\ref{sec:adversaries} discusses adversaries who may be motivated and capable to conduct the attacks discussed in section~\ref{sec:https_scenarios}.

\subsection{Privacy Applications of HTTPS}
\label{sec:https_scenarios}
We present several categories of website in which the specific pages accessed by the user are more interesting than the mere fact that the user is visiting the website at all.
This notion is present in traditional privacy concepts such as patient confidentiality or attorney-client privilege, where the content of a communication is substantially more sensitive than the simple presence of communication.

\textbf{Healthcare}
Many medical conditions or procedures are associated with significant social stigma.
We examine the websites of Planned Parenthood, Mayo Clinic and Kaiser Permanente, a healthcare provider serving 9 million members in the US.
The page views of these websites have the potential to reveal whether a pending procedure is an appendectomy or an abortion, or whether a chronic medication is for diabetes or HIV/AIDS.
These types of distinctions and others can form the basis for discrimination or persecution and represent an easy opportunity to target advertising for products which consumers are highly motivated to purchase.
Beyond personal risks, the health care details of corporate and political leaders can also have significant financial implications, as evidenced by Apple stock fluctuations in response to reports, both true and false, of Steve Jobs's health~\cite{jobs_health}.

\textbf{Legal}
There are many common reasons for interaction with a lawyer, such as completing a will, filing taxes, or reviewing a contract.
However, contacting a lawyer to investigate divorce, bankruptcy, or legal options as an undocumented immigrant may attract greater interest.
Since some legal advice is relatively unremarkable while other advice may require strict privacy, the specific details of legal services are more interesting than mere interaction with a lawyer.
Our work examines LegalZoom, a website offering legal services spanning the above themes and others.
We additionally examine the American Civil Liberties Union (ACLU), which offers legal information and actively litigates on a wide range of sensitive topics including LGBT rights, human reproduction and immigration.

\textbf{Financial}
While most consumers utilize some form of financial products to manage their personal finances, the exact products a person uses reveal a great deal more about their personal circumstances.  
For example, a user with educational savings accounts likely has children, a joint account is an indicator of a long term relationship, and high volume mutual funds offering reduced fees likely indicate high levels of minimum net worth.
Our work examines Bank of America and Wells Fargo, both large banks in the US, as well as Vanguard, a firm offering a range of investment vehicles and brokerage services.

\textbf{Streaming Video}
As demonstrated during the Netflix Prize contest and ensuing \$9 million settlement, the video rental history of an individual can potentially reveal information as personal as sexual orientation~\cite{netflix_lawsuit,netflix_settlement}.
Beyond any guarantees given in privacy policies, video rentals in the US are additionally protected by law~\cite{vppa}.
We examine YouTube and Netflix, both of which offer streaming videos covering a wide range of topics.

\subsection{Attack Settings}
\label{sec:adversaries}
Having reviewed the possible consequences of traffic analysis attacks against HTTPS, we now examine situations in which an adversary may be motivated and capable to learn the types of private details previously discussed.
Note that all capable adversaries must have at least two abilities.
The adversary must be able to visit the same webpages as the victim, allowing the adversary to identify patterns in encrypted traffic indicative of different webpages.
The adversary must also be able to observe victim traffic, allowing the adversary to match observed traffic with previously learned patterns.

\textbf{ISP Snooping}
ISPs are uniquely well positioned to target and sell advertising since they have the most comprehensive view of the consumer.
Both ISPs~\cite{nyt,register} and commercial chains of wi-fi access points~\cite{wifi_attack}, have shown efforts to mine customer data and/or sell advertising.
Traffic analysis vulnerabilities would allow ISPs to conduct data mining despite the presence of encryption.
Separate from electronic ad delivery, access points associated with businesses such as cafes and hotels could also deliver ads along with transaction receipts, physical mailings, or other special offers.

\textbf{Employee Monitoring}
Employers have the ability to monitor the online activities of employees connected to an employer provided network, regardless of whether the device in use is a personal or corporate device.
This power has been abused by extensively monitoring the activities of employees~\cite{workplace}, even extending to whistleblowers whose communications are protected by law~\cite{whistleblowers}.
Traffic analysis would allow employers to remove many of the protections expected by employees using HTTPS to protect their sensitive communications from untrusted parties.

\textbf{Surveillance}
While revelations of NSA surveillance spanning from social media to World of Warcraft are an unwelcome surprise to many~\cite{wow_snoop,lover_snoop,prism,xkeyscore}, other governments around the world have long employed these practices~\cite{china_censorship,iran_censorship}.
When asked about the efficacy of encryption, Snowden maintained  ``Encryption works. Properly implemented strong crypto systems are one of the few things that you can rely on. Unfortunately, endpoint security is so terrifically weak that NSA can frequently find ways around it''~\cite{encryption_works}.
Despite this assertion, we still see NSA surveillance efforts specifically targeting HTTPS~\cite{bullrun}, indicating the value of removing side-channel attacks to ensure that HTTPS is ``properly implemented.''

\textbf{Censorship}
Although the consequences of forbidden internet activity can include imprisonment and beyond in some settings, in other contexts broad filtering efforts have resulted in lower grade punishments designed to deter further transgression and encourage self-censorship.
For example, Chinese social media firm Sina has recently punished more than 100,000 users through account suspensions and occasional public admonishment for violating the country's ``Seven Bottom Lines'' guidelines for internet use~\cite{weibo}.
Similarly, traffic analysis attacks could be used to degrade or block service for users suspected of viewing prohibited content over encrypted connections.

\begin{table*}[t]
\begin{center}
\resizebox{\textwidth}{!}{
\begin{tabular}[width=\textwidth]{l l l l l l l l l l}
								&Privacy 		&Page Set\;\;		&Page Set\;\; 	&Accuracy	&			& 			&Traffic			&Analysis 		& Active		\\
Author							&Technology	&Scope			&Size		&(\%)		&Cache\;\;		& Cookies		&Composition\;\;	&Primitive\;\;	& Content		\\ \hline

\textbf{Miller}						&\textbf{HTTPS}&\textbf{Closed}	&\textbf{6396}	&\textbf{89}	&\textbf{On}	& \textbf{Individual}\;\;	&\textbf{Single Site}	&\textbf{Packet}	&\textbf{On} \\ \hline
Hintz~\cite{hintz}					&SSL proxy	&Closed			&5			&100			&?			& Individual	&Homepages		&Request		& ?	  \\ 
\multirow{2}{*}{Sun~\cite{sun}}			&\multirow{2}{*}{SSL proxy}	&\multirow{2}{*}{Open}			&2,000		&75 (TP)		&\multirow{2}{*}{Off}	&\multirow{2}{*}{Universal}	&\multirow{2}{*}{Single Site}	&\multirow{2}{*}{Request}	& \multirow{2}{*}{Off}	\\
								&						&							&100,000		&1.5	(FP)		& 				&						&				&					&				\\ \hline

Cheng~\cite{cheng}					&HTTPS		&Closed			&489			&96			&Off			& Individual	&Single Site		&Request		& Off		\\ 
Danezis~\cite{danezis} 				&HTTPS		&Closed			&?			&89			&n/a			& n/a			&Single Site		&Request		& n/a 	\\ \hline

Herrmann~\cite{herrmann}			&SSH tunnel	&Closed			&775			&97			&Off			& Universal	&Homepages		&Packet		& ?	  		\\
Cai~\cite{cai}						&SSH tunnel	&Closed			&100			&92			&Off			& Universal	&Homepages		&Packet		& Scripts		\\
Dyer~\cite{dyer}					&SSH tunnel	&Closed			&775			&91			&Off			& Universal	&Homepages		&Packet		& ?	  		\\
Liberatore	~\cite{liberatore}\;\;			&SSH tunnel\;\;	&Closed			&1000		&75			&Off			& Universal	&Homepages		&Packet		& Flash		\\
Bissias~\cite{bissias}					&SSH tunnel	&Closed			&100			&23			&?			& Universal	&Homepages		&Packet		& ?	  		\\ \hline

\multirow{2}{*}{Wang~\cite{wang}}		&\multirow{2}{*}{Tor}	&\multirow{2}{*}{Open}	&100			&95 (TP)		&\multirow{2}{*}{Off}	& \multirow{2}{*}{Universal}	&\multirow{2}{*}{Homepages}	&\multirow{2}{*}{Packet}	&\multirow{2}{*}{Off}\\
								&				&					&1000		&.06 (FP)		&				&						&						&					&				\\
Wang~\cite{wang}					&Tor			&Closed			&100			&91			&Off			& Universal	&Homepages		&Packet	& Off				\\
Cai~\cite{cai}						&Tor			&Closed			&100			&78			&On			& Universal	&Homepages		&Packet	& Scripts		\\
Cai~\cite{cai}						&Tor			&Closed			&800			&70			&Off			& Universal	&Homepages		&Packet	& Scripts		\\
Panchenko~\cite{panchenko}			&Tor			&Closed			&775			&55			&Off			& Universal	&Homepages		&Packet	& Off				\\
\multirow{2}{*}{Panchenko~\cite{panchenko}}	&\multirow{2}{*}{Tor}	&\multirow{2}{*}{Open}	&5 			&56-73 (TP)		&\multirow{2}{*}{Off}	& \multirow{2}{*}{Universal}	&\multirow{2}{*}{Homepages}		&\multirow{2}{*}{Packet}	&\multirow{2}{*}{Off}	\\
									&				&					&1,000		&.05-.89	(FP)\;\;	& 				&						&							&					&				\\
Herrmann~\cite{herrmann}			&Tor			&Closed			&775			&3			&Off			& Universal	&Homepages		&Packet	&?	\\ \hline

\multirow{2}{*}{Coull~\cite{coull}}		&Anonymous	&\multirow{2}{*}{Open}	&50			&49			&\multirow{2}{*}{On}	&\multirow{2}{*}{Universal}	&\multirow{2}{*}{Homepages}	&\multirow{2}{*}{NetFlow}	& Flash \& \\
								&Trace		&					&100			&.18			&				&						&						&					& Scripts \\
\end{tabular}
}
\end{center}
\small
\caption{Prior works have focused almost exclusively on website homepages accessed via proxy.  Cheng and Danezis work is preliminary and unpublished.  Evaluations for both works parse object sizes from unencrypted traffic or server logs, which is not possible for actual encrypted traffic. Note that ``?'' indicates the author did not specify the property; several properties did not apply to Danezis as his evaluation used HTTP server logs.  All evaluations used Linux with Firefox (FF) 2.0-3.6, except for Hintz and Sun (IE5), Cheng (Netscape), Wang (FF10) and Miller (FF22).}
\label{table:rel_work}
\end{table*}

\section{Prior Work}
\label{sec:rel_work}

In this section we review attacks and defenses proposed in prior work, as well as the contexts in which work is evaluated.
Comparisons with prior work are limited since much work has targeted specialized technologies such as Tor.

Table~\ref{table:rel_work} presents an overview of prior attacks.
The columns are as follows:

\begin{description}
\item[Privacy Technology] The encryption or protection mechanism analyzed for traffic analysis vulnerability.  Note that some authors considered multiple mechanisms, and hence appear twice.

\item[Page Set Scope] \textit{Closed} indicates the evaluation used a fixed set of pages known to the attacker in advance.  \textit{Open} indicates the evaluation used traffic from pages both of interest and unknown to the attacker.  Whereas open conditions are appropriate for Tor, closed conditions are appropriate for HTTPS.

\item[Page Set Size] For closed scope, the number of pages used in the evaluation.  For open scope, the number of pages of interest to the attacker and the number of background traffic pages, respectively.

\item[Accuracy] For closed scope, the percent of pages correctly identified.  For open scope, the true positive (TP) rate of correctly identifying a page as being within the censored set and false positive (FP) rate of identifying an uncensored page as censored.

\item[Cache] \textit{Off} indicates caching disabled. \textit{On} indicates default caching behavior.

\item[Cookies] \textit{Universal} indicates that training and evaluation data were collected on the same machine or machines, and consequently with the same cookie values.  \textit{Individual} indicates training and evaluation data were collected on separate machines with distinct cookie values.

\item[Traffic Composition] \textit{Single Site} indicates the work identified pages within a website or websites.  \textit{Homepages} indicates all pages used in the evaluation were the homepages of different websites.

\item[Analysis Primitive] The basic unit on which traffic analysis was conducted.  \textit{Request} indicates the analysis operated on the size of each object (e.g. image, style sheet, etc.) loaded for each page.  \textit{Packet} indicates meta-data observed from TCP packets.  \textit{NetFlow} indicates network traces anonymized using NetFlow.

\item[Active Content] Indicates whether Flash, JavaScript, Java or any other plugins were enabled in the browser.

\end{description}

Several works require discussion in addition to Table~\ref{table:rel_work}.
Danezis focused on the HTTPS context, but evaluated his technique using HTTP server logs at request granularity, removing any effects of fragmentation, concurrent connections or pipelined requests~\cite{danezis}.
Cheng \textit{et al.} also focused on HTTPS and conducted an evaluation using traffic from an HTTP website intentionally selected for its static content~\cite{cheng}.
Both works were unpublished, and operated on individual object sizes parsed from the unencrypted traffic rather than packet metadata.
Likewise, the approaches of Sun~\textit{et al.} and Hintz~\textit{et al.} also assume the ability to parse entire object sizes from traffic~\cite{hintz,sun}.
For these reasons, we compare our work to Liberatore and Levine, Panchenko~\textit{et al.} and Wang~\textit{et al.} as these are more advanced and recently proposed techniques.

Herrmann~\cite{herrmann} and Cai~\cite{cai} both conduct small scale preliminary evaluations which involve enabling the browser cache.
In contrast to our evaluation, these evaluations only consider website homepages and all pages are loaded in a fixed, round-robin order.
Herrmann additionally increases the cache size from the default to 2GB, reducing the likelihood of any cache evictions and stabilizing traffic.
With caching enabled, Herrmann and Cai both observe approximately a 5\% decrease in accuracy for their techniques, and Cai reports slightly improved performance for the Panchenko classifier.
We evaluate the impact of caching on pages within the same website, where caching will have a greater effect than on the homepages of different websites due to increased page similarity, and load pages in a randomized order for greater cache state variation.

Separate from attacks, we also review prior work relating to traffic analysis defense.
Dyer~\textit{et al.} conduct a review of low level defenses operating on individual packets~\cite{dyer}.
Dyer evaluates defenses using data released by Liberatore and Levine and Herrmann~\textit{et al.} which collect traffic from website home pages on a single machine with caching disabled.
In this context, Dyer finds that low level defenses are ineffective against attacks which examine features aggregated over multiple packets.
For example, the linear and exponential padding defenses, which pad packet sizes to multiples of 128 and powers of 2 respectively, reduce the accuracy of the Panchenko classifier at most from 97.2\% to 96.6\%.
In our evaluation, which considers pages within the same website, enabled caching and identification of traces collected on machines separate from the attacker, we find that low level, stateless defenses can be considerably more effective than initially indicated by Dyer.

In addition to the packet level defenses evaluated by Dyer, many defenses have been proposed which operate at higher levels with additional cost and implementation requirements.
These include HTTPOS~\cite{luo}, traffic morphing~\cite{wright} and BuFLO~\cite{cai,dyer}.
HTTPOS, unlike most defenses, works from the client side to perturb the traffic generated by manipulating various features of TCP and HTTP to affect packet size, object size, pipelining behavior, packet timing and other properties.
These manipulations require some degree of coordination and support from the server.
BuFLO aims to provide provable defense against traffic analysis attacks by sending a constant stream of traffic at a fixed packet size for a pre-set minimum amount of time.
Given the effectiveness and advantages of lower level defenses in our evaluation context, we do not further explore these higher level approaches in our work.

\section{Attack Presentation}
\label{sec:attack_presentation}
\label{sec:approach}
\begin{figure*}[t]
        \begin{center}
        	        \includegraphics[width=\textwidth]{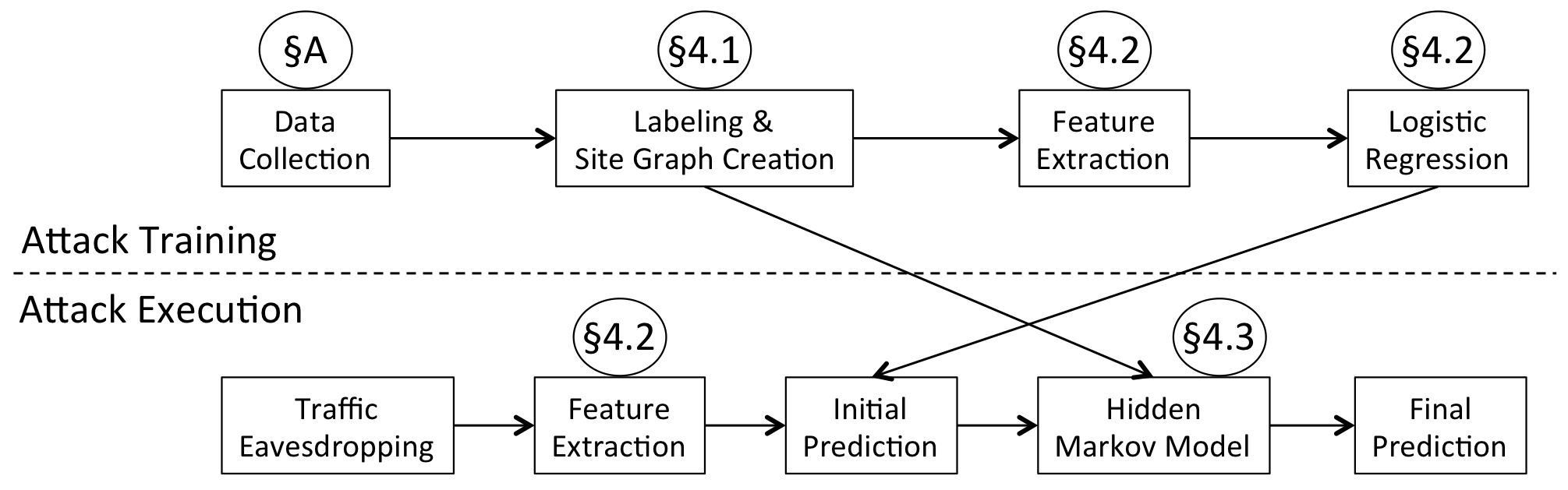}
        \end{center}
        \caption{Attack Presentation. The dashed line separates training workflow (above) from attack workflow (below). Bubbles indicate the section in which the system component is discussed, with \S A indicating Appendix A.}
        \label{fig:system_diagram}
\end{figure*}

In this section we present our attack.
Figure~\ref{fig:system_diagram} presents an overview of the attack, depicting the anticipated workflow of the attacker as well as the subsections in which we discuss his efforts.
In section~\ref{sec:naming}, we present a formalism for identifying and labeling pages within a website and 
generating a site graph representing the website link structure.
Section~\ref{sec:gaussians} presents the core of our classification approach: Gaussian clustering techniques that capture 
standard variations in traffic and allow logistic regression to robustly identify key objects which reliably differentiate pages.
Having generated isolated predictions, we then leverage the site graph and sequential nature of the data in section~\ref{sec:hmm} with a hidden Markov model (HMM) to further improve accuracy.

Throughout this section we depend on several terms which we define as follows:
\begin{description}
\item[Uniform Resource Locator (URL)] A character string referencing a specific web resource, such as an image or HTML file.
\item[Webpage] The set of resources loaded by a browser in response to the user clicking a link or otherwise entering a URL into the browser address bar.  Two webpages are \textit{the same} if a user could be reasonably expected to view their contents as substantially similar, regardless of the specific URLs fetched while loading the webpages or dynamic content such as advertising.
\item[Sample] An instance of the traffic generated when a browser displays a webpage.
\item[Label] A unique identifier assigned to each set of webpages which are the same.  For example, two webpages which differ only in advertising will receive the same label, while webpages differing in core content are assigned different labels.  Labels are assigned to samples in accordance with the webpage contained in the sample's traffic.
\item[Website] A set of webpages such that the URLs which cause each webpage to load in the browser are hosted at the same domain.
\item[Site Graph] A graph representing the link structure of a website. Nodes correspond to labels assigned to webpages within the website.  Edges correspond to links between webpages, represented as the set of labels reachable from a given label.
\end{description}

\subsection{Label and Site Graph Generation}
\label{sec:naming}
This section presents our approach to labeling and site graph generation.
Merely treating the URL which causes a webpage to load as the label for the webpage is not sufficient for analyzing webpages within the same website.
URLs may contain arguments, such as session IDs, which do not impact the content of the webpage and result in different labels aliasing webpages which are the same.
This prevents accumulation of sufficient training samples for each label and hinders evaluation.
URL redirection further complicates labeling; the same URL may refer to multiple webpages (e.g. error pages or A/B testing) or multiple URLs may refer to the same webpage.
We present a labeling solution based on URLs and designed to accommodate these challenges.
\footnote{While URL redirection may be implemented within the web server or via JavaScript that alters webpage contents, allowing a single URL to represent many webpages, this behavior is limited in practice because website designers are motivated to allow search engines to link to webpages in search results.}
When labeling errors are inevitable, we prefer to have a single webpage aliased to multiple labels rather than have multiple distinct pages aliased to a single label.
The former may result in lower accuracy ratings, but it allows our attacker to learn \textit{correct} information.

Our approach contains two phases.
In the first phase, we conduct a preliminary crawl of the website, yielding many URLs from links encountered during the crawl.
We then analyze these URLs to produce a \textit{canonicalization function} which, given a URL, returns a canonical label for the webpage loaded as result of entering the URL into a browser address bar.\footnote{Note that we label pages based on the final URL after any URL redirection occurs.}
We use the canonicalization function to produce a preliminary site graph, which guides further crawling activity.
Our approach proceeds in two phases because the non-deterministic nature of URL redirections requires the attacker to conduct extensive crawling to observe the full breadth of both URLs and redirections, and crawling can not be conducted without a basic heuristic identifying URLs which likely alias the same webpage.
As we describe below our approach allows, but does not require, the second crawl to be combined with training data collection.
After the second phase is complete, both the labels and site graph are refined using the additional URLs and redirections observed during the crawl.
We present our approach below.

\textbf{Execute Preliminary Crawl}
The first step in developing labels and a site graph is to crawl the website.
The crawl can be implemented as either a depth- or breadth-first search, beginning at the homepage and exploring every link on a page up to a fixed maximum depth.
We perform a breadth first search to depth 5.
This crawl will produce a graph $G = (U, E)$, where $U$ represents the set of URLs seen as links during the crawl, and $E = \{(u, u') \in U \times U \; | \; u$ links to $u'\}$ represents links between URLs in $U$.

\textbf{Produce Canonicalization Function}
\label{sec:canonicalize}
Since multiple URLs may cause webpages which are effectively the same to load when entered into a browser address bar, the role of a canonicalization function is to produce a canonical label given a URL.
The canonicalization function will be of the form $C: U \rightarrow L$, where $C$ denotes the canonicalization function, $U$ denotes the initial set of URLs, and $L$ denotes the set of labels.
To maintain the criterion that we error on the side of multiple labels aliasing the same webpage, our approach forces any URLs with different paths to be assigned different labels and selectively identifies URL arguments that appreciably impact webpage content for inclusion in the label.
We were able to execute this phase on all websites we surveyed.
See Appendix B for our full approach, including several heuristics independent of URL arguments which further guide canonicalization.

\textbf{Canonicalize Initial Graph}
We use our canonicalization function to produce an initial site graph $G' = (L, E')$ where $L$ represents the set of labels on the website and $E'$ represents links.
We construct $E'$ as follows:
\begin{eqnarray}
E' = \{(C(u), C(u')) \: | \: (u, u') \in E\}
\end{eqnarray}
We define a reverse canonicalization function $R: L \rightarrow \mathcal P(U)$ such that
\begin{equation}
R(l) = \{u \in U \: | \: C(u) = l\}
\end{equation}
Note that $\mathcal P(X)$ denotes the \textit{power set} of $X$, which is the set of all subsets of $X$.

\textbf{Identify Browsing Sessions}
The non-deterministic nature of URL redirection requires the attacker to observe many redirection examples to finalize the site graph and canonicalization function.
This process can also be used to collect training data.
To collect training data and observe URL redirections the attacker builds a list of browsing sessions, each consisting of a fixed length sequence of labels.
We fix the length of our browsing sessions to 75 labels.
For cache accuracy, the attacker builds browsing sessions using a random walk through $G'$.
Since the graph structure prevents visiting all nodes evenly, the attacker prioritizes labels not yet visited.
When the portion of duplicate labels reaches a fixed threshold (we used 0.6), the attacker visits the remaining labels regardless of the graph link structure until all labels have received at least single visit.
This process is repeated until the attacker has produced enough browsing sessions to collect the desired amount of training data; we collected at least 64 samples of each label in total.

\textbf{Execute Browsing Sessions}
To generate traffic samples the attacker selects a URL $u$ for each label $l$ in a browsing session such that $u \in R(l)$ and loads $u$ and (all supporting resources) by effectively entering $u$ into a browser address bar.
The attacker records the value of \texttt{document.location} (once the entire webpage is done loading) to identify any URL redirections.
$U'$ denotes the set of final URLs which are observed in \texttt{document.location}.
We define a new function $T: U \rightarrow \mathcal P(U')$ such that $T(u) = \{ u' \in U' \; | \; u$ resolved at least once to $u'\}$.
We use this to define a new translation $T': L \rightarrow  \mathcal P(U')$ such that
\begin{equation}
T'(l) = \bigcup_{u \in R(l)} T(u)
\end{equation}

\textbf{Refine Canonicalization Function}
Since the set of final URLs $U'$ may include arguments which were not present in the original set $U$, we refine our canonicalization function $C$ to produce a new function $C': U' \rightarrow L'$, where $L'$ denotes a new set of labels.
The refinement is conducted using the same techniques as we used to produced $C$.
Samples are labeled as $C'(u') \in L'$ where $u'$ denotes the value of \texttt{document.location} when the sample finished loading.

\textbf{Refine Site Graph}
Since the final set of labels $L'$ may contain labels which are not in $L$, the attacker must update $G'$.
The update must maintain the property that any sequence of labels $l'_0, l'_1, ... \in L'$ observed during data collection must be a valid path in the final graph.
Therefore, the attacker defines a new graph $G'' = (U', E'')$ such that 
$E''$ is defined as
\begin{equation}
E''  = \{(u, u') \: | \: u \in T'(l) \: \wedge \: u' \in T'(l') \: \forall \: (l, l') \in E' \}
\end{equation}
We apply our canonicalization function $C'$ to produce a final graph $G''' = (L', E''')$ where
\begin{equation}
E''' = \{(C'(u), C'(u')) \: \forall \: (u, u') \in E''\}
\end{equation}
maintaining the property that any sequence of labels observed during training is a valid path in the final graph.
Note that our evaluation generates a separate site graph for each model, using only redirections which occurred in training data.
This leaves the possibility of a path in evaluation data which is not valid on the attacker site graph, but we did not find this to be an issue in practice.

\begin{table*}[t]
\begin{center}
\resizebox{\textwidth}{!}{
\begin{tabular}[width=.95\textwidth]{ l | r r r | r r r | r r}
						& \multicolumn{3}{c|}{Preliminary Site Graph} & \multicolumn{3}{c|}{Selected Subset} & \multicolumn{2}{c}{Final Site Graph}\\
Website		 			& \,URLs\,		& \,Labels\,		& \, Avg. Links\,	& \,URLs\,	& \,Labels\,& \,Avg. Links\, & \,Labels\, & \,Avg. Links\\ \hline
ACLU &	54398 &	28383 &	130.5 &	1061 &	500 &	41.7 & 506 & 44.7\\
Bank of America &	1561 &	613 &	30.2 &	1105 &	500 &	30.3 & 472 & 43.2\\
Kaiser Permanente &	1756 &	780 &	29.7 &	1030 &	500 &	22.6  & 1037 & 141.1\\
Legal Zoom &	5248 &	3973 &	26.8 &	602 &	500 &	11.8 & 485 & 12.2\\
Mayo Clinic &	33664 &	33094 &	38.1 &	531 &	500 &	12.5 & 990 & 31.0\\
Netflix &	8190 &	5059 &	13.8 &	2938 &	500 &	6.2 & 926 & 9.0\\
Planned Parenthood &	6336 &	5520 &	29.9 &	662 &	500 &	24.8 & 476 & 24.4\\
Vanguard &	1261 &	557 &	28.4 &	1054 &	500 &	26.7 & 512 & 30.8\\
Wells Fargo &	4652 &	3677 &	31.2 &	864 &	500 &	17.9 & 495 & 19.5\\
YouTube &	64348 &	34985 &	7.9 &	953 &	500 &	4.3 & 497 & 4.24\\
\end{tabular}
}
\end{center}
\caption{Site graph and canonicalization summary.  ``Selected Subset'' denotes the subset of the preliminary site graph which we randomly select for inclusion in our evaluation, ``Avg. Links'' denotes the average number of links per label, and ``URLs'' indicates the number of URLs seen as links in the preliminary site graph corresponding to an included label.}
\label{table:graph_transformation}
\end{table*}

For the purposes of this work, we augment the above approach to select a subset of the preliminary site graph for further analysis.
By surveying a subset of each website, we are able to explore additional websites and browser configurations and remain within our resource constraints.
We initialize the selected subset to include the label corresponding to the homepage, and iteratively expand the subset by adding a label reachable from the selected subset via the link structure of the preliminary site graph until 500 labels are selected.
The set of links for the graph subset is defined as any links occurring between the 500 selected labels.
Table~\ref{table:graph_transformation} presents properties of the preliminary site graph $G'$, selected subset, and the final site graph $G'''$ for each of the 10 websites we survey.
We implement the preliminary crawl using Python and the second crawl (i.e. training data collection) using the browsing infrastructure described in Appendix A.

\subsection{Feature Extraction and Machine Learning}
\label{sec:gaussians}

This section presents our individual sample classification technique.
First, we describe the information which we extract from a sample, then we describe processing to produce features for machine learning, and finally describe the application of the learning technique itself.
\begin{figure*}[t]
        \begin{center}
		\subfloat[]{\label{fig:dom_a}\includegraphics[width=0.45\textwidth]{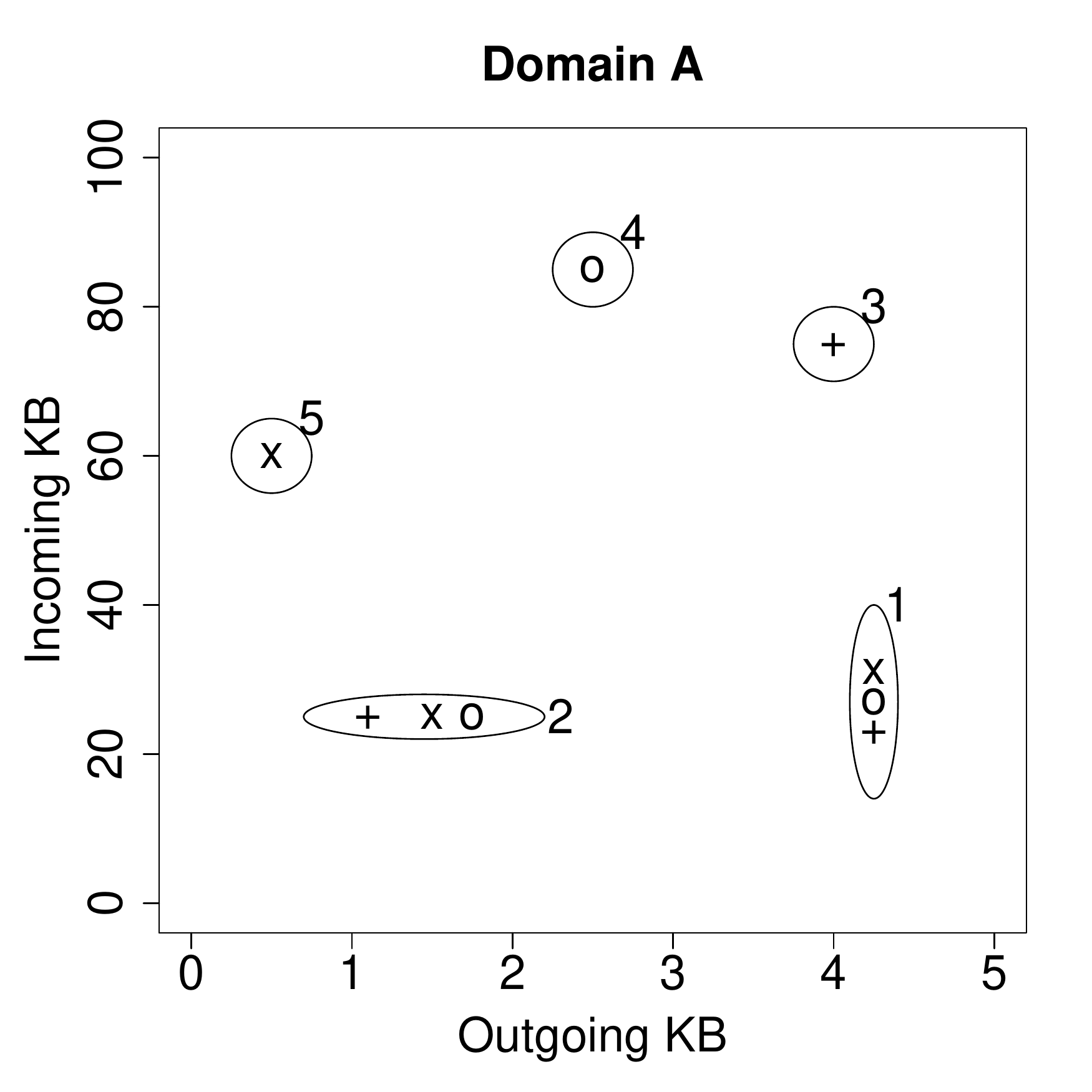}}
        	        \subfloat[]{\label{fig:dom_b}\includegraphics[width=0.45\textwidth]{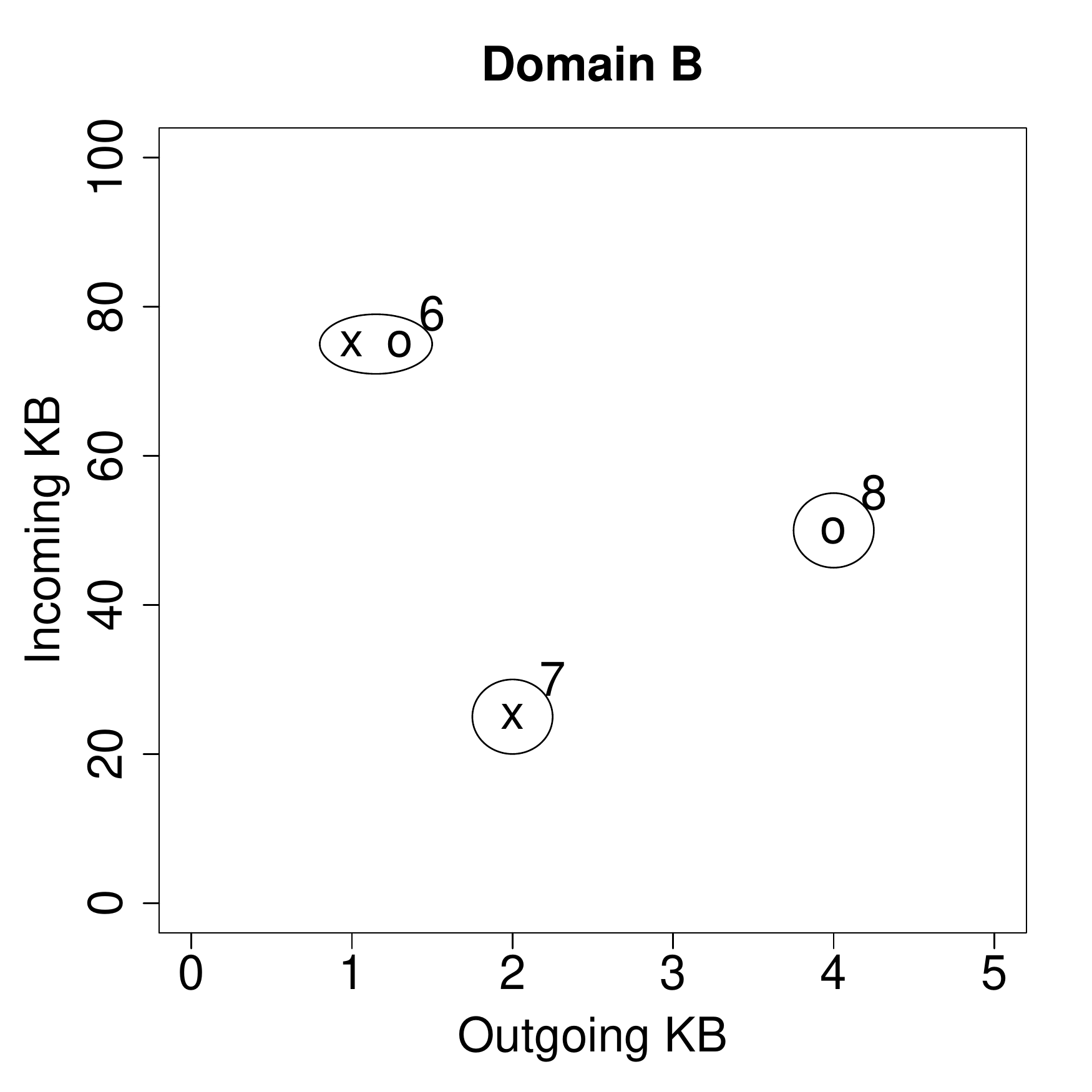}}
	        \end{center}
	\vspace{-.7cm}
	        	\begin{center}
	        \subfloat[]{
\begin{tabular}[b]{l | c c c | c c }
\multicolumn{6}{c}{\textbf{Burst Pairs (KB)}} \\
Domain & & A & & \multicolumn{2}{ c }{B} \\ \hline
Sample \texttt{x} & (0.5, 60) & (1.6, 22) & (4.2, 30) & (1.1, 75) & (2.0, 25) \\
Sample \texttt{+} & (1.1, 22) & (4.0, 75) & (4.2, 21) &  & \\
Sample \texttt{o} & (1.8, 22) & (2.4, 83) & (4.2, 25) & (1.4, 75) & (4.0, 50) \\
\end{tabular}
\label{fig:burst_pairs}
}
	        \subfloat[]{
\begin{tabular}[b]{l | c c c c c | c c c}
\multicolumn{9}{c}{\textbf{Feature Values}} \\
Domain & & & A & & & & B & \\ \hline
Index &1 & 2 & 3 & 4 & 5 & 6 & 7 & 8 \\ \hline
Sample \texttt{x} & .9 & 1 & 0 & 0 & 1 & .8 & 1 & 0 \\
Sample \texttt{+} & .9 & .8 &1 & 0 & 0 & 0 & 0 & 0 \\
Sample \texttt{o} & 1 & .9 & 0 & 1 & 0 & .8 & 0 & 1 \\
\end{tabular}
\label{fig:bog_features}
}
        \end{center}
	\vspace{-.2cm}
       \caption{Table~\ref{fig:burst_pairs} displays the burst pairs extracted from three hypothetical samples.  Figures~\ref{fig:dom_a} and~\ref{fig:dom_b} show the result of burst pair clustering. Figure~\ref{fig:bog_features} depicts the Bag-of-Gaussians features for each sample, where each feature value is defined as the total likelihood of all points from a given sample at the relevant domain under the Gaussian distribution corresponding to the feature.  Our Gaussian similarity metric enables our attack to distinguish minor traffic variations from significant differences.}
	\vspace{-.2cm}
        \label{fig:clustering}
\end{figure*}

We initially extract traffic burst pairs from each sample.
Burst pairs are defined as the collective size of a contiguous outgoing packet sequence followed by the collective size of a contiguous incoming packet sequence.
Intuitively, contiguous outgoing sequences correspond to requests, and contiguous incoming sequences correspond to responses.
All packets must occur on the same TCP connection to minimize the effects of interleaving traffic.
For example denoting outgoing packets as positive and incoming packets as negative, the sequence \texttt{[+1420, +310, -1420, -810, +530, -1080]} would result in the burst pairs \texttt{[1730, 2230]} and \texttt{[530, 1080]}.
Analyzing traffic bursts removes any fragmentation effects.
Additionally, treating bursts as pairs allows the data to contain minimal ordering information and go beyond techniques which focus purely on packet size distributions.

Once burst pairs are extracted from each TCP connection, the pairs are grouped using the second level domain of the host associated with the destination IP of the connection.
All IPs for which the reverse DNS lookup fails are treated as a single ``unknown'' domain.
Pairs from each domain undergo k-means clustering to identify commonly occurring and closely related tuples.
Since tuples correspond to individual requests and pipelined series of requests, some tuple values will occur on multiple webpages while other tuples will occur only on individual webpages.
Once clusters are formed we fit a Gaussian distribution to each cluster and treat each cluster as a feature dimension, producing our fixed-width feature vector.
Features are extracted from samples by computing the extent to which each Gaussian is represented in the sample.

Figure~\ref{fig:clustering} depicts the feature extraction process using a fabricated example involving three samples and two domains.
Clustering results in five clusters, indexed 1--5, for Domain A and three clusters, indexed 6--8, for Domain B.
The feature vector thus has eight dimensions, with one corresponding to each cluster.
Sample \texttt{x} has 
traffic tuples in clusters 1, 2, 5, 6 and 7, but no traffic tuples in clusters 3, 4, 8, so its feature
vector has non-zero values in dimensions 1, 2, 5, 6, 7, and zero values in dimensions 3, 4, 8. 
We create feature vectors for samples \texttt{+} and \texttt{o} in a similar fashion.

We specify our approach formally as follows:
\begin{itemize}
\item Let $X$ denote the entire set of tuples from a trace, with $X^d \subseteq X$ denoting set all tuples observed at domain $d$.
\item Let $\Sigma^d_i, \mu^d_i$ respectively denote the covariance and mean of Gaussian $i$ at domain $d$.
\item Let $F$ denote all features, with $F^d_i$ denoting feature $i$ from domain $d$.
\end{itemize}
\begin{equation}
F^d_i = \sum_{x \in X^d} \mathcal{N}(x | \Sigma^d_i, \mu^d_i)
\end{equation}
To determine the best number of Gaussian features for each domain, we train models using a range of values of $K$ and then select the best performing model for each domain.

Analogously to the Bag-of-Words document processing technique, our approach projects a variable length vector of tuples into a finite dimensional space where each dimension ``occurs'' to some extent in the original sample.
Whereas occurrence is determined by word count in Bag-of-Words, occurrence in our method is determined by Gaussian likelihood.
For this reason, we refer to our approach as Bag-of-Gaussians (BoG).

Once Gaussian features have been extracted from each sample the feature set is augmented to include counts of packet sizes observed over the entire trace.
For example, if the lengths of all outgoing and incoming packets are between 1 and 1500 bytes, we add 3000 additional features where each feature corresponds to the total number of packets sent in a specific direction with a specific size.
We linearly normalize all features to be in the range $[0, 1]$ and train a model using L2 regularized multi-class logistic regression with $C = 128$ using the \texttt{liblinear} package~\cite{liblinear}.

\subsection{Hidden Markov Model}
\label{sec:hmm}

The basic attack presented in section~\ref{sec:gaussians} classifies each sample independently.
In practice, samples in a browsing session are not independent since the link structure of the website guides the browsing sequence.
We leverage this ordering information, contained in the site graph produced in section~\ref{sec:naming}, to improve results using a hidden Markov model (HMM).
Recall that a HMM for a sequence of length $N$ is defined by a set of latent variables $Z = \{z_n \; | \; 1 \leq n \leq N\}$, a set of observed variables $X = \{x_n \; | \; 1 \leq n \leq N\}$, transition matrix $A$ such that $A_{i,j} = P(Z_{n+1} = j | Z_n = i)$, an initial distribution $\pi$ such that $\pi_j = P(Z_1 = j)$ and an emission function $E(x_n, z_n) = P(x_n | z_n)$.

Applied to our context, the HMM is configured as follows:
\begin{itemize}
\item Latent variables $z_n$ correspond to labels $l' \in L'$ visited by the victim during browsing sessions
\item Observed variables $x_n$ correspond to observed feature vectors $X$
\item Initial distribution $\pi$ assigns an equal likelihood to all pages
\item Transition matrix $A$ encodes $E'''$, the set of links between pages in $L'$, such that all links have equal likelihood
\item Emission function $E(x_n, z_n) = P(z_n | x_n)$ determined by logistic regression
\end{itemize}

After obtaining predictions with logistic regression, the attacker refines the predictions using the Viterbi algorithm to solve for the most likely values of the latent variables, each of which corresponds to a pageview by the user.

\section{Impact of Evaluation Conditions}
\label{sec:eval}

In this section we demonstrate the impact of evaluation conditions on accuracy results and traffic characteristics.
First, we present the scope and motivation of our investigation.
Then, we describe the experimental methodology we use to determine the impact of evaluation conditions.
Finally, we present the results of our experiments on four attack implementations.
All attacks are impacted by the evaluation condition variations we consider, with the most affected attack decreasing accuracy from 68\% to 50\%.
We discuss attack accuracy in this section only insofar as is necessary to understand the impact of evaluation conditions; we defer a full attack evaluation to section~\ref{sec:attack_eval}.

\textbf{Cache Configuration}
The browser cache improves page loading speed by storing web resources which were previously loaded, potentially posing two challenges to traffic analysis.
Providing content locally decreases the total amount of traffic, reducing the information available for use in an attack.
Additionally, differences in browsing history can result in differences in cache contents and further vary network traffic.
Since privacy tools such as Tor frequently disable caching, many prior evaluations have disabled caching as well~\cite{tor}.
But in practice, general users of HTTPS typically do not modify default cache settings, so we evaluate the impact of enabling caching to default settings.

\textbf{User-Specific Cookies}
If an evaluation collects all data with either the same browser instance or repeatedly uses a fresh browser image (such as the Tor browser bundle), there are respective assumptions that the attacker and victim will either share the same cookies or use none at all.
While a traffic analysis attacker will not have access to victim cookies, privacy technologies which begin all browsing sessions from a clean browsing image effectively share the \textit{null cookie}.
We compare the performance of evaluations which use the same (non-null) cookie value in all data, different (non-null) cookie values in training and evaluation, a null cookie in all data, and evaluations which mix null and non-null cookies.

\textbf{Pageview Diversity}
Many evaluations collect data by repeatedly visiting a fixed set of URLs from a single machine and dividing the collected data for training and evaluation. 
This approach introduces an unrealistic assumption that, during training, an attacker will be able to visit the same set of webpages as the victim.
Note that this would require collecting separate training data for each victim given that each victim visits a unique set of pages.
We examine the impact of allowing the victim to intersperse browsing of multiple websites, including websites outside our attacker's monitoring focus.\footnote{Note that this is different from an open-world vs. closed-world distinction as described in section~\ref{sec:rel_work}, as we assume that the attacker will train a model for each website in its entirety and be able to identify the correct model based on traffic destination.  Here, we are concerned with any effects on browser cache or personalized content which may impact traffic analysis.}

\textbf{Webpage Similarity}
Since HTTPS will usually allow an eavesdropper to learn the domain a user is visiting, our evaluation focuses on efforts to differentiate individual webpages within a website protected by HTTPS.
Differentiating webpages within the same website may pose a greater challenge than differentiating website homepages.
Webpages within the same website share many resources, increasing the effect of caching and making webpages within a website harder to distinguish.
We examine the relative traffic volumes of browsing both website homepages and webpages within a website.

To quantify the impact of evaluation conditions on accuracy results, we design four modes for data collection designed to isolate specific effects.
Our approach assumes that data will be gathered in a series of browsing sessions, where each session consists of loading a fixed number of URLs in a browser.
The four modes are as follows:
\begin{enumerate}
\item Cache disabled, new virtual machine (VM) for each browsing session
\item Cache enabled, new VM for each browsing session
\item Cache enabled, persistent VM for all browsing sessions, single site per VM
\item Cache enabled, persistent VM for all browsing sessions, all sites on same VM
\end{enumerate}

\begin{figure*}[t]
	\begin{center}
	\includegraphics[width=.95\textwidth]{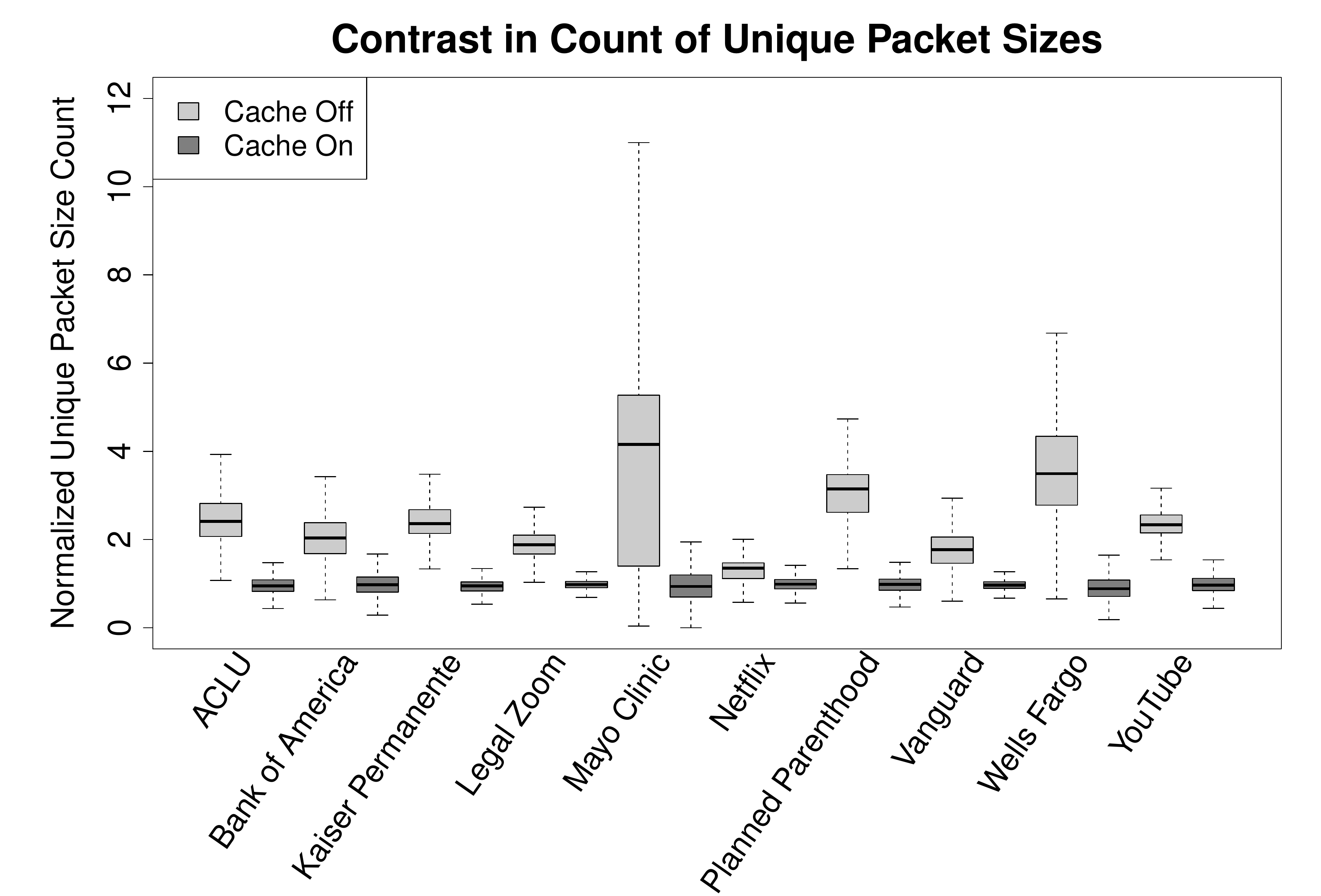}
	\end{center}
	\caption{Disabling the cache significantly increases the number of unique packet sizes for samples of a given label.  For each label $l$ we determine the mean number $l_m$ of unique packet sizes for samples of $l$ with caching enabled, and normalize the unique packet size counts of all samples of label $l$ using $l_m$.  We present the normalized values for all labels separated by cache configuration.}
	\label{fig:size_count}
\end{figure*}

In our experiments we fixed the session length to 75 URLs and collect at least 16 samples of each label under each collection mode.
We begin each browsing session in the first two configurations with a fresh VM image to eliminate the possibility of cookie persistence in browser or machine state.
The first and second modes differ only with respect to cache configuration, allowing us to quantify the impact of caching.
In effect the second, third and fourth modes each represent a distinct cookie value, with the second mode representing a null cookie and the third and fourth modes having actual, distinct, cookie values set by the site.
The third and fourth modes differ in pageview diversity.
In the context of HTTPS evaluations, the fourth mode most closely reflects the behavior of actual users and hence serves as evaluation data, while the second and third modes generate training data.

\begin{figure*}[t]
        \begin{center}
       	        \subfloat[]{\label{fig:cache_compare}\includegraphics[width=.266\textwidth]{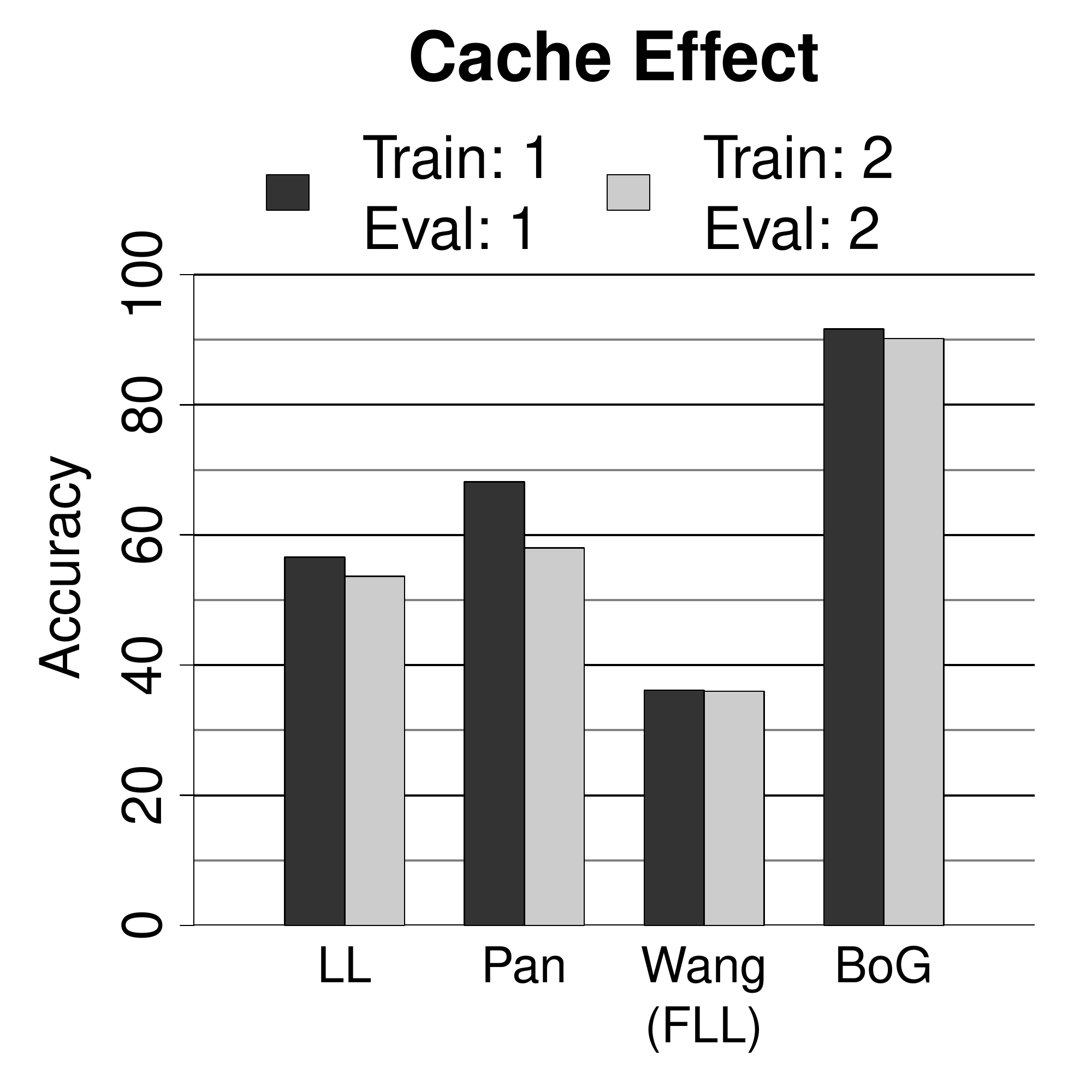}}        
       	        \subfloat[]{\label{fig:cookie_presence}\includegraphics[width=.398\textwidth]{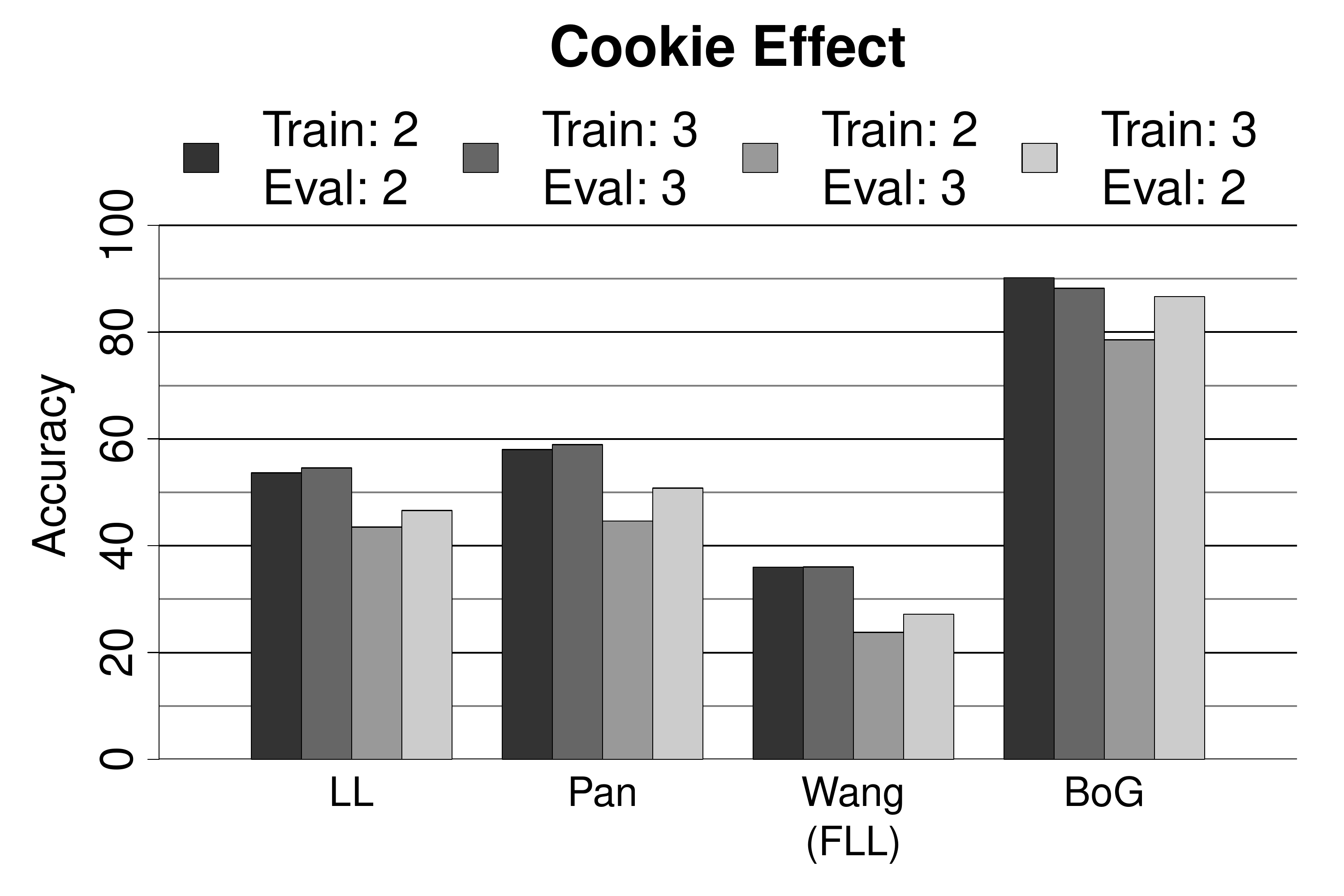}}        
       	        \subfloat[]{\label{fig:net_effect}\includegraphics[width=.333\textwidth]{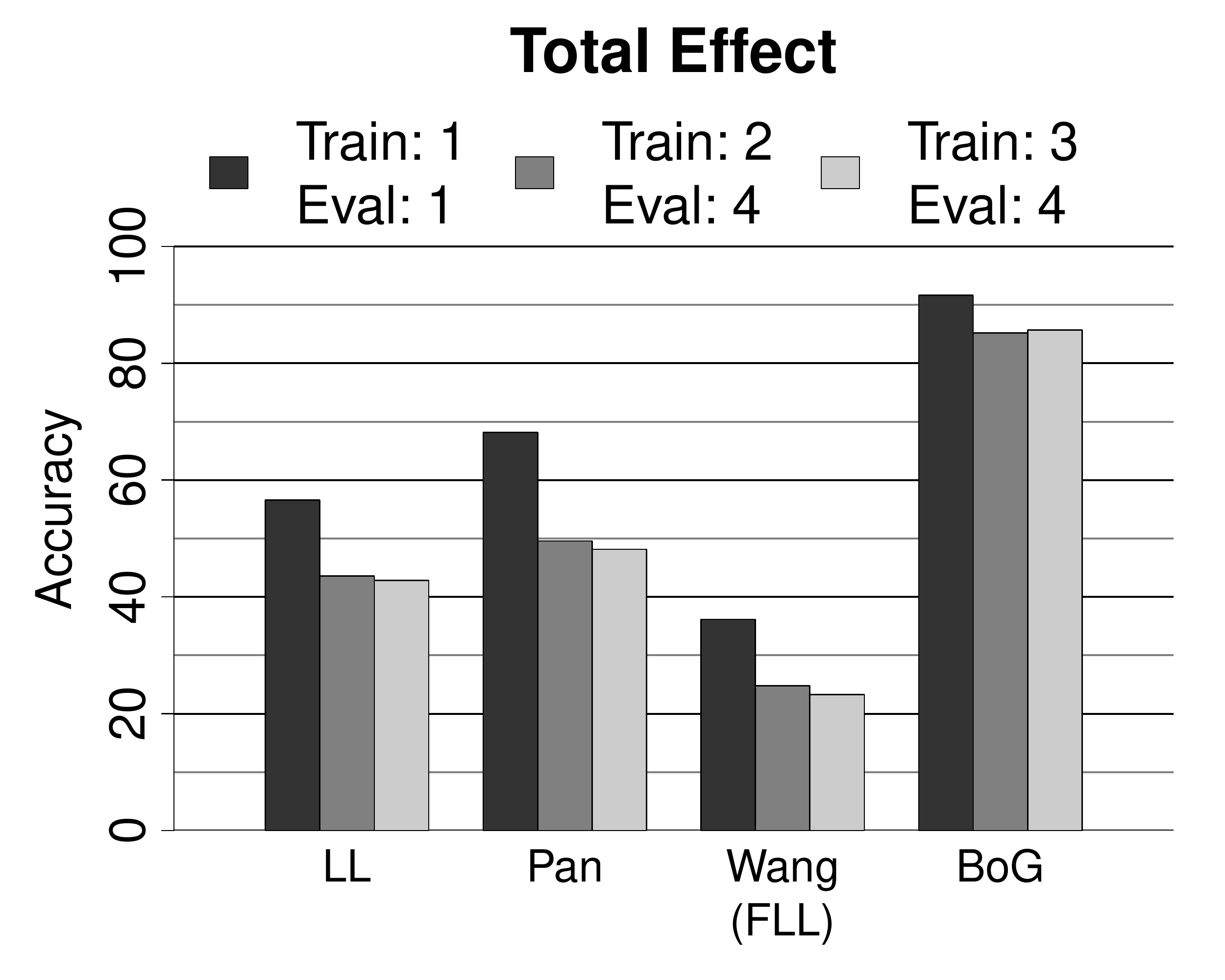}}        	        
        \end{center}
        	\begin{center}
	        \subfloat[]{
\begin{tabular}[b,width=\textwidth]{ c | c |  c | c }
Mode Number\;  	& \;Cache\;  	&  \;Cookie Retention\; 	&  \;Browsing Scope \\ \hline
1			& \;Disabled\;	& \;Fresh VM every 75 samples\;			&  Single website \\
2			& Enabled		& Fresh VM every 75 samples			&  Single website \\
3			& Enabled		& Same VM for all samples			&  Single website \\
4			& Enabled		& Same VM for all samples		&  All websites \\
\end{tabular}
\label{fig:legend}
}
	\end{center}
        \caption{``Train: X, Eval: Y'' indicates training data from mode X and evaluation data from mode Y as shown in Table~\ref{fig:legend}.  For evaluations which use a privacy tool such as the Tor browser bundle and assume a closed world, training and evaluating using mode 1 is most realistic.  However, in the HTTPS context training using mode 2 or 3 and evaluating using mode 4 is most realistic. Figure~\ref{fig:net_effect} presents differences as large as 18\% between these conditions, demonstrating the importance of evaluation conditions when measuring attack accuracy.}
        	\label{fig:eval_conditions}
\end{figure*}

Our analysis reveals that caching significantly decreases the number of unique packet sizes observed for samples of a given label.
We focus on the number of unique packet sizes since packet size counts are a commonly used feature in traffic analysis attacks.
A reduction in the number of unique packet sizes reduces the number of non-zero features and creates difficulty in distinguishing samples.
Figure~\ref{fig:size_count} contrasts samples from the first and second collection modes, presenting the effect of caching on the number of unique packet sizes observed per label for each of the 10 websites we evaluate.
Note that the figure only reflects TCP data packets.
We use a normalization process to present average impact of caching on a \textit{per-label basis} across an \textit{entire website}, allowing us to depict for each website the expected change in number of unique packet sizes for any given label as a result of disabling the cache.

\begin{figure*}[t]
	\begin{center}
	\includegraphics[width=.6\textwidth]{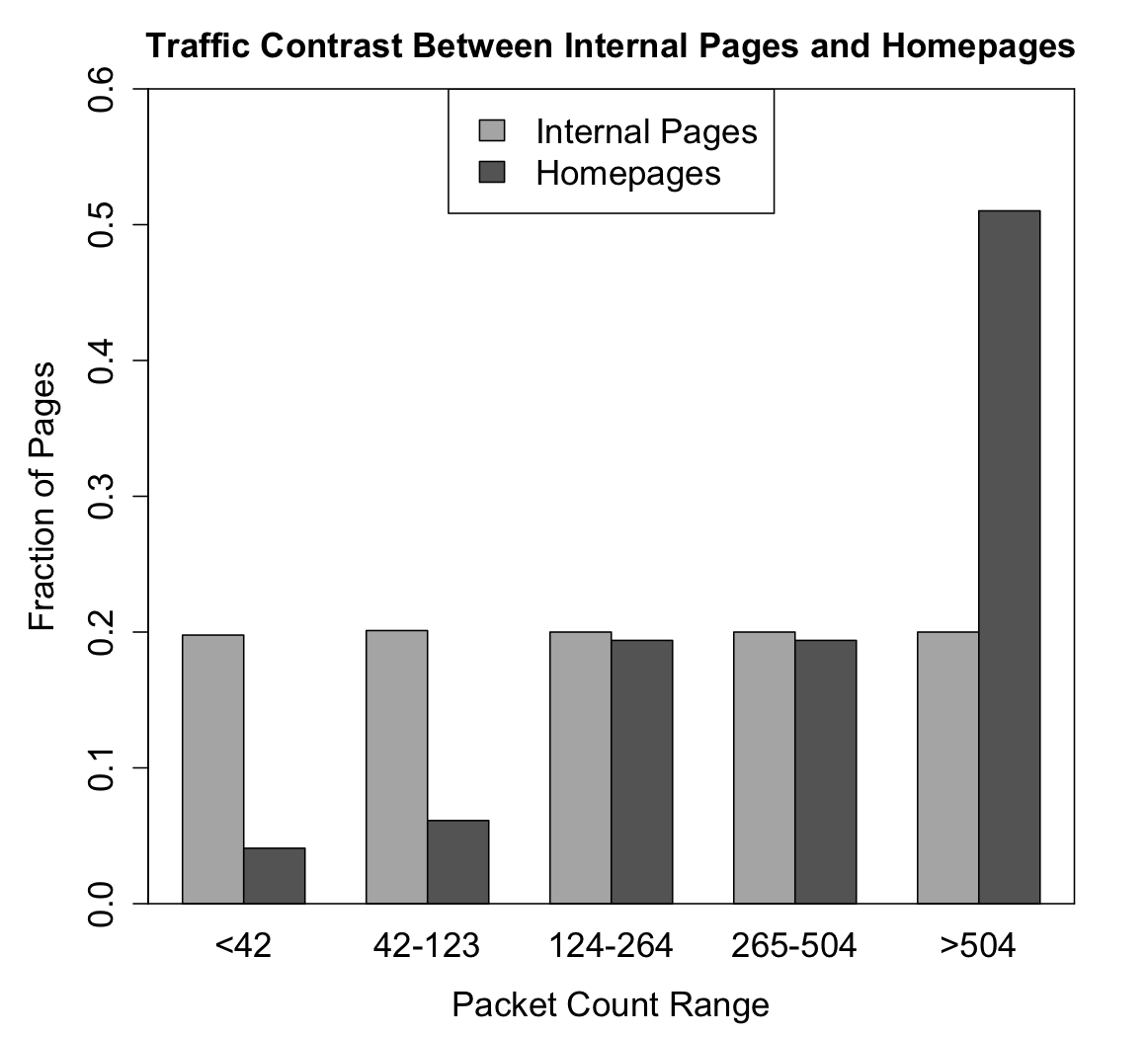}
	\end{center}
        \caption{Decrease in traffic volume caused by browsing webpages internal to a website as compared to website homepages.  Similar to the effect of caching, the decreased traffic volume is likely to increase classification errors. Note that packet count ranges are selected to divide internal pages into 5 even ranges.}
	\label{fig:packet_count}
\end{figure*}

Beyond examining traffic characteristics, our analysis shows that factors such as caching, user-specific cookies and pageview diversity can cause variations as large as 18\% in measured attack accuracy.
We examine each of these factors by training a model using data from a specific collection mode, and comparing model accuracy when evaluated on a range of collection modes.
Since some models must be trained and evaluated using data from the same collection mode, we use 8 training samples per label and leave the remaining 8 samples for evaluation.
Figure~\ref{fig:eval_conditions} presents the results of our analysis:
\begin{description}
\item[Cache Effect] Figure~\ref{fig:cache_compare} compares the performance of models trained and evaluated using mode 1 to models trained and evaluated using mode 2.
As these modes differ only by enabled caching, we see that caching has moderate impact and can influence reported accuracy by as much as 10\%.
\item[Cookie Effect] Figure~\ref{fig:cookie_presence} measures the impact of user-specific cookies by comparing the performance of models trained and evaluated using browsing modes 2 and 3.
We observe that both the null cookie in mode 2 and the user-specific cookie in mode 3 perform 5-10 percentage points better when the evaluation data is drawn from the same mode as the training data. 
This suggests that any difference in cookies between training and evaluation conditions will impact accuracy results.
\item[Total Effect] Figure~\ref{fig:net_effect} presents the combined effects of enabled caching, user-specific cookies and increased pageview diversity.
Recalling Figure~\ref{fig:cookie_presence}, notice that models trained using mode 2 perform similarly on modes 3 and 4, and models trained using mode 3 perform similarly on modes 2 and 4, confirming the importance of user-specific cookies.
In total, the combined effect of enabled caching, user-specific cookies and pageview diversity can influence reported accuracy by as much as 19\%.
Figure~\ref{fig:cookie_presence} suggests that the effect is primarily due to caching and cookies since mode 2 generally performs better on mode 4, which includes visits to other websites, than on mode 3, which contains traffic from only a single website.
\end{description}

Since prior works have focused largely on website homepages but HTTPS requires identification of webpages within the same website, we present data demonstrating a decrease in traffic when browsing webpages within a website.
Figure~\ref{fig:packet_count} presents the results of browsing through the Alexa top 1,000 websites, loading the homepage of each site, and then loading nine additional links on the site at random with caching enabled.
By partitioning the total count of data packets transferred in the loading of webpages internal to a website into five equal size buckets, we see that there is a clear skew towards homepages generating more traffic, reflecting increased content and material for traffic analysis.
This increase, similar to the increase caused by disabled caching, is likely to increase classification errors.

\section{Attack Evaluation}
\label{sec:attack_eval}

In this section we evaluate the performance of our attack.
We begin by presenting the selection of previous techniques for comparison and the implementation of each attack.
Then, we present the aggregate performance of each attack across all 10 websites we consider, the impact of training data on attack accuracy, and the performance each attack at each individual website.

We select the Liberatore and Levine (LL), Panchenko \textit{et al.} (Pan), and Wang \textit{et al.} attacks for evaluation in addition to the BoG attack.
The LL attack offers a view of baseline performance achievable from low level packet inspection, applying naive Bayes to a feature set consisting of packet size counts~\cite{liberatore}.
We implemented the LL attack using the naive Bayes implementation in \texttt{scikit-learn}~\cite{scikit-learn}.
The Pan attack extends size count features to include additional features related to burst length as measured in both packets and bytes as well as total traffic volume~\cite{panchenko}.
For features aggregated over multiple packets, the Pan attack rounds feature values to predetermined intervals.
We implement the Pan attack using the \texttt{libsvm}~\cite{libsvm} implementation of the RBF kernel support vector machine with the $C$ and $\gamma$ parameters specified by Panchenko.
We select the Pan attack for comparison to demonstrate the significant benefit of Gaussian similarity rather than predetermined rounding thresholds.
The BoG attack functions as described in section~\ref{sec:approach}.
We implement the BoG attack using the k-means package from \texttt{sofia-ml}~\cite{sofia-ml} and logistic regression with class probability output from \texttt{liblinear}~\cite{liblinear}, with \texttt{Numpy}~\cite{numpy} performing intermediate computation.

The Wang attack assumes a fundamentally different approach from LL, Pan and BoG based on string edit distance~\cite{wang}.
There are several variants of the Wang attack which trade computational cost for accuracy by varying the string edit distance function.
Wang reports that the best distance function for raw packet traces is the Optimal String Alignment Distance (OSAD) originally proposed by Cai~\textit{et al.}~\cite{cai}.
Unfortunately, the edit distance must be computed for each pair of samples, and OSAD is extremely expensive.
Therefore, we implement the Fast Levenshtein-Like (FLL) distance,\footnote{Note that the original attack rounded packet sizes to multiples of 600; we operate on raw packet sizes as we found this improves attack accuracy in our evaluation.} 
an alternate edit distance function proposed by Wang which runs approximately 3000x faster.\footnote{OSAD has $O(mn)$ runtime where $m$ and $n$ represent the length of the strings, whereas FLL runs in $O(m+n)$.
Wang \textit{et al.} report completing an evaluation with 40 samples of 100 labels each in approximately 7 days of CPU time.
Since our evaluation involves 10 websites with approx. 500 distinct labels each and 16 samples of each label for training and evaluation, we would require approximately 19 months of CPU time (excluding any computation for sections~\ref{sec:eval} or~\ref{sec:defense_eval}).}
Since Wang demonstrates that FLL achieves 46\% accuracy operating on raw packet traces, as compared to 74\% accuracy with OSAD, we view FLL as a rough indicator of the potential of the OSAD attack.
We implement the Wang - FLL attack using \texttt{scikit-learn}~\cite{scikit-learn}.

We now examine the performance of each attack implementation.
We evaluate attacks using data collected in mode 4 since this mode is most similar to the behavior of actual users.
We consider both modes 2 and 3 for training data to avoid any bias introduced by using the same cookies as seen in evaluation data or browsing the exact same websites.
As shown in Figure~\ref{fig:eval_conditions}, mode 2 performs slightly better so we train all models using data from mode 2.

\begin{figure*}[t]
        \begin{center}
        \subfloat[]{\label{fig:train_impact}\includegraphics[width=.49\textwidth]{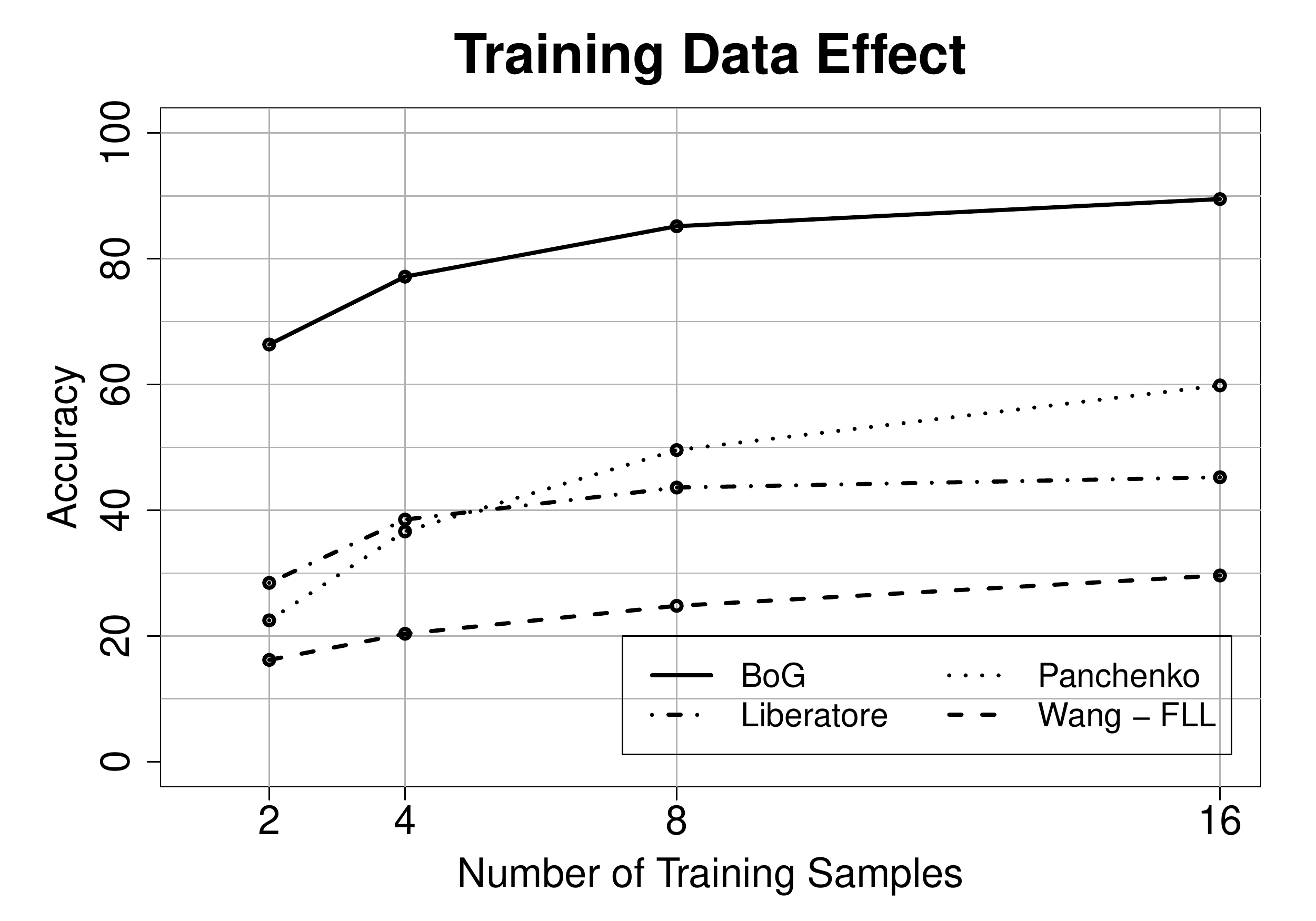}}
        \subfloat[]{\label{fig:session_length_impact}\includegraphics[width=.49\textwidth]{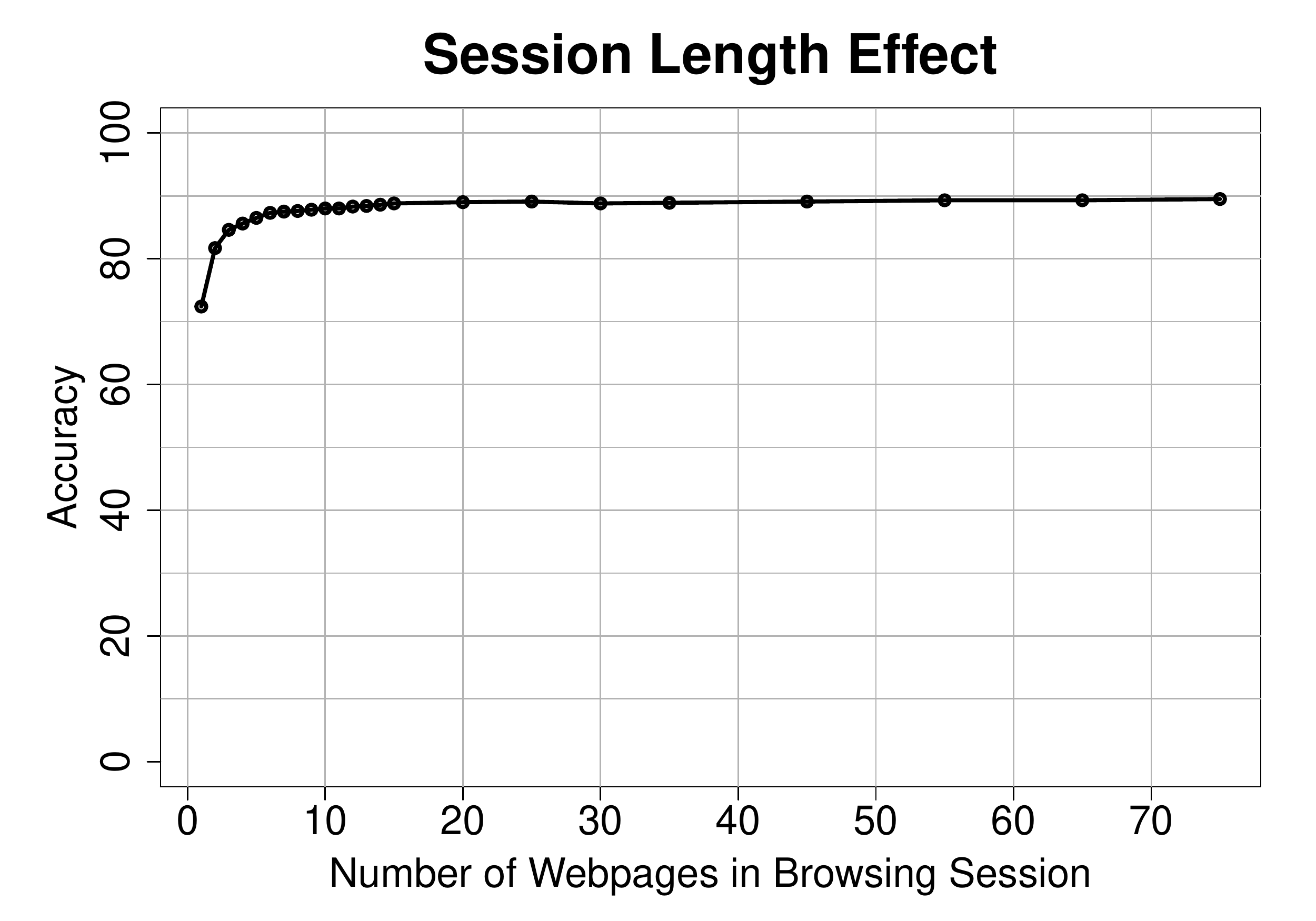}}
        \end{center}
        \vspace{-.5cm}
	\caption{Performance of BoG attack and prior techniques.  Figure~\ref{fig:train_impact} presents the performance of all four attacks as a function of training data.  Figure~\ref{fig:session_length_impact} presents the accuracy of the BoG attack trained with 16 samples as a function of browsing session length.  Note that the BoG attack achieves 89\% accuracy as compared to 60\% accuracy with the best prior work.}
        \vspace{-.5cm}
        \label{fig:accuracy_overview}
\end{figure*}

Consistent with prior work, our evaluation finds that accuracy of each attack improves with increased training data, as indicated by Figure~\ref{fig:train_impact}.
Notice that the Pan attack is most sensitive to variations in the amount of training data, and the BoG attack continues to perform well even at low levels of training data.
In some cases an attacker may have ample opportunity to collect training data, although in other cases the victim website may attempt to actively resist traffic analysis attacks by detecting crawling behavior and engaging in cloaking, rate limiting or other forms of blocking.
These defenses would be particularly feasible for special interest, low volume websites where organized, frequent crawling would be hard to conceal.

The BoG attack derives significant performance gains from the application of the HMM.
Figure~\ref{fig:session_length_impact} presents the BoG attack accuracy as a function of the browsing session length.
Although we collect browsing sessions which each contain 75 samples, we simulate shorter browsing sessions by applying the HMM to randomly selected subsets of browsing sessions and observing impact on accuracy.
At session length 1 the HMM has no effect and the BoG attack achieves 72\% accuracy, representing the improvement over the Pan attack resulting from the Gaussian feature extraction.
The HMM accounts for the remaining performance improvement from 72\% accuracy to 89\% accuracy.
We achieve most of the benefit of the HMM after observing two samples in succession, and the full benefit after observing approximately 15 samples.
Although any technique which assigns a likelihood to each label for each sample could be extended with a HMM, applying a HMM requires solving the labeling and site graph challenges which we present novel solutions for in section~\ref{sec:approach}.

\begin{figure*}[]
        \begin{center}
        \includegraphics[width=\textwidth]{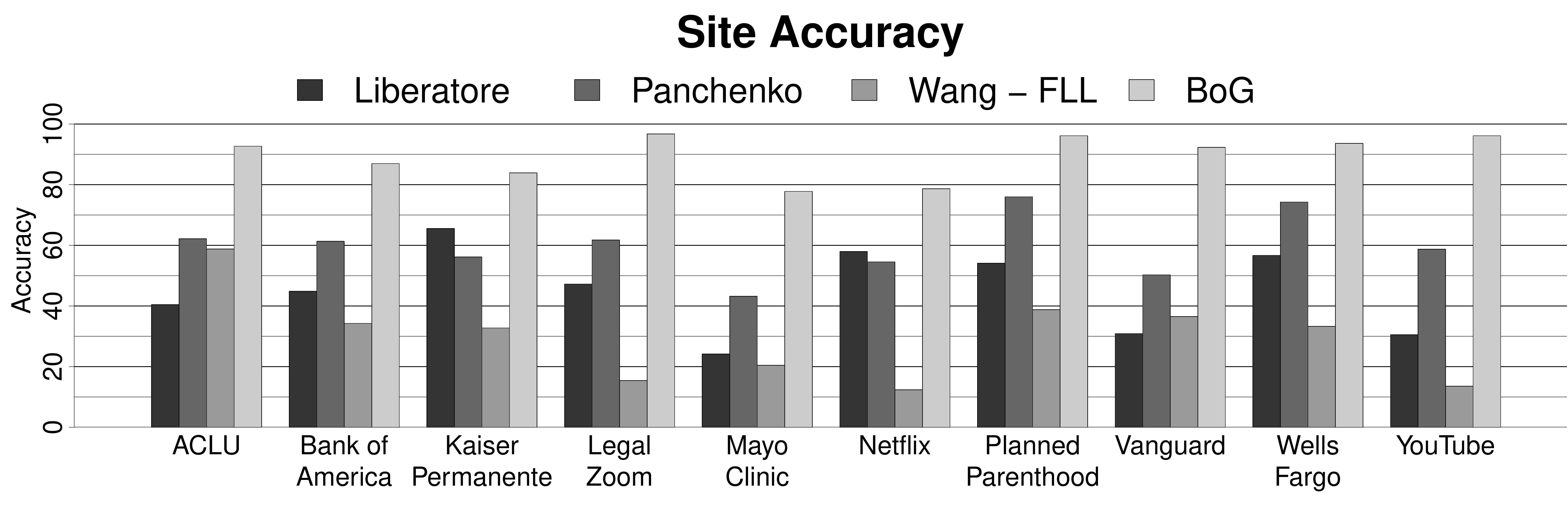}
        \end{center}
        \vspace{-.5cm}
        \caption{Accuracy of each attack for each website.  Note that the BoG attack performs the worst at Kaiser Permanente, Mayo Clinic and Netflix, which each have approx. 1000 labels in their final site graphs according to Table~\ref{table:graph_transformation}. The increase in graph size during finalization suggests potential for improved performance through better canonicalization to eliminate labels aliasing the same webpages.}
        \vspace{-.5cm}
        \label{fig:site_accuracy}
\end{figure*}

Although the BoG attack averages 89\% accuracy overall, only 4 of the 10 websites included in evaluation have accuracy below 92\%.
Figure~\ref{fig:site_accuracy} presents the accuracy of each attack at each website.
The BoG attack performs the worst at Mayo Clinic, Netflix and Kaiser Permanente.
Notably, the number of labels in the site graphs corresponding to each of these websites approximately doubles during the finalization process summarized in Table~\ref{table:graph_transformation} of section~\ref{sec:approach}.
URL redirection causes the increase in labels, as new URLs appear whose corresponding labels were not included in the preliminary site graph.
Some new URLs may have been poorly handled by the canonicalization function, resulting in labels which alias the same content.
Although we collected supplemental data to gather sufficient training samples for each label, the increase in number of labels and label aliasing behavior degrade measured accuracy for all attacks.

Despite the success of string edit distance based attacks against Tor, the Wang - FLL attack struggles in the HTTPS setting.
Wang's evaluation is confined to Tor, which pads all packets into fixed size cells, and does not effectively explore edit distance approaches applied to unpadded traffic.
Consistent with the unpadded nature of HTTPS, we observe that Wang's attack performs best on unpadded traffic in the HTTPS setting.
Despite this improvement, the Wang - FLL technique may struggle because edit distance treats all unique packet sizes as equally dissimilar; for example, 310 byte packets are equally similar to 320 byte packets and 970 byte packets.
Additionally, the application of edit distance at the packet level causes large objects sent in multiple packets to have proportionally large impact on edit distance.
This bias may be more apparent in the HTTPS context than with website homepages since webpages within the same website are more similar than homepages of different websites.
Replacing the FLL distance metric with OSAD or Damerau-Levenshtein would improve attack accuracy, although the poor performance of FLL suggests the improvement would not justify the cost given the alternative techniques available.

\section{Defense}
\label{sec:defense_eval}

\begin{figure*}[t]
        \begin{center}
		\subfloat[]{\label{fig:low_level_on}\includegraphics[width=.6\textwidth]{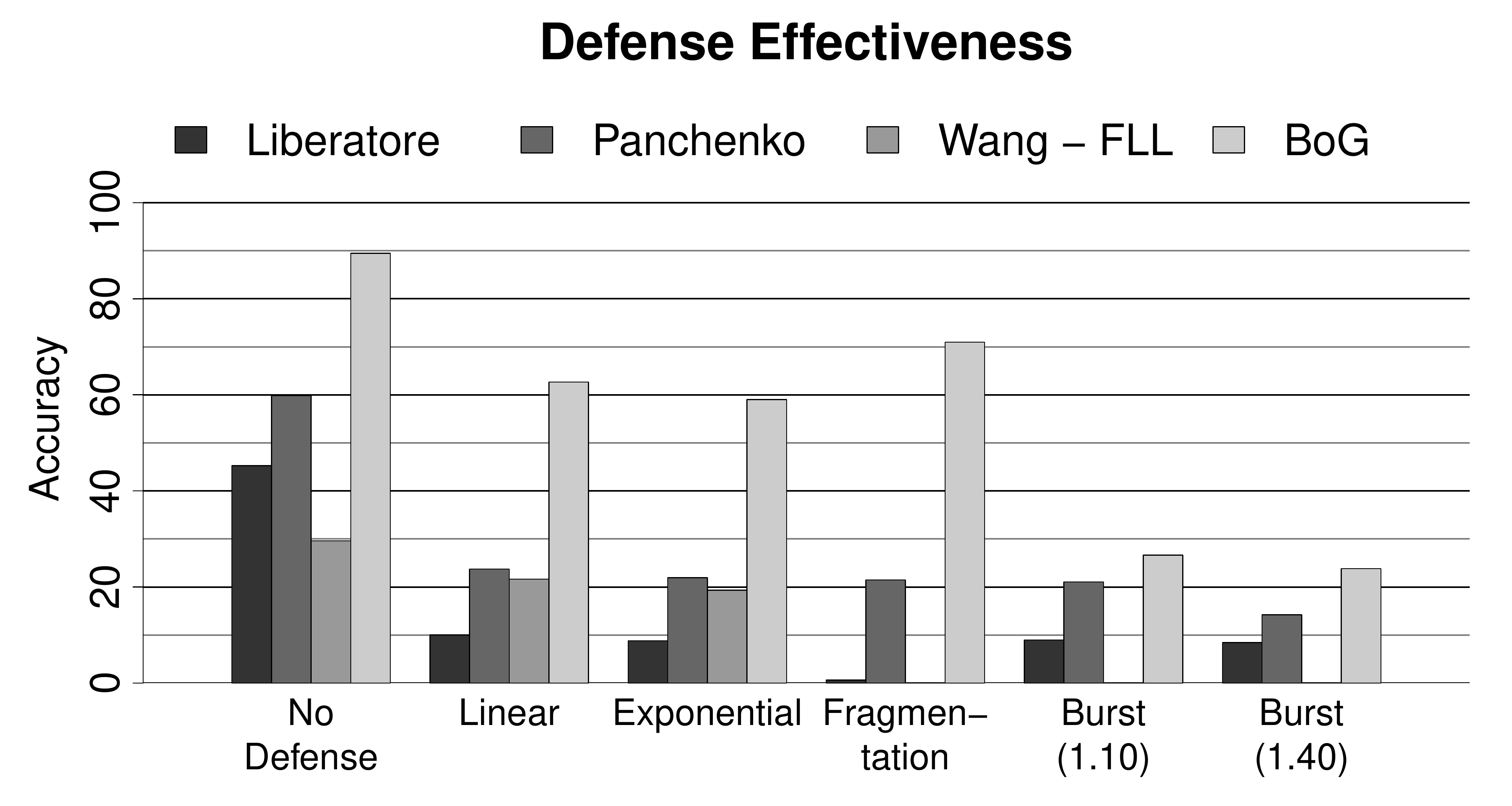}}
        	        \subfloat[]{\label{fig:cost}

\resizebox{.38\textwidth}{!}{
	        \begin{tabular}[b,width=\textwidth]{ l | c | c}
 \multicolumn{3}{c}{ \vspace{-5cm} }\\
		& Byte		& Packet   \\
Defense 	& Overhead 	& Overhead \\ \hline
None &	1.000 &	 1.000  \\
Linear &	1.032 &	 1.000 \\
Exponential &	1.055 &	 1.000 \\
Fragmentation &	1.000 &	 1.450 \\
Burst: 1.10 &	1.090 &	 1.065 \\
Burst: 1.40 &	1.365 &	 1.248 \\
  \end{tabular} 
  }
	        }
        \end{center}
       \caption{Cost and effectiveness of defense techniques.  Figure~\ref{fig:low_level_on} presents the impact of defenses on attack accuracy, and Figure~\ref{fig:cost} presents the cost of defenses.  The Burst defense is a novel proposal in this work, offering a substantial decrease in accuracy at a cost comparable to a low level defense.}
        \label{fig:def_performance}
\end{figure*}

This section presents and evaluates several defense techniques, including our novel Burst defense which operates between the application and TCP layers to obscure high level features of traffic while minimizing overhead.
Figure~\ref{fig:def_performance} presents the effectiveness and cost of the defenses we consider.
We find that evaluation context significantly impacts defense performance, as we observe increased effectiveness of low level defenses in our evaluation as compared to prior work~\cite{dyer}.
Additionally, we find that the Burst defense offers significant performance improvements while maintaining many advantages of low level defense techniques.

We select defenses for evaluation on the combined basis of cost, deployability and effectiveness.
We select the linear and exponential padding defenses from Dyer \textit{et al.} as they are reasonably effective, have the lowest overhead, and are implemented statelessly below the TCP layer.
The linear defense pads all packet sizes up to multiples of $128$, and the exponential defense pads all packet sizes up to powers of $2$.
Stateless implementation at the IP layer allows for easy adoption across a wide range of client and server software stacks.
Additionally, network overhead is limited to minor increases in packet size with no new packets generated, keeping costs low in the network and on end hosts.
We also introduce the fragmentation defense which randomly splits any packet which is smaller than the MTU, similar to portions of the strategy adopted by HTTPOS~\cite{luo}.
Fragmentation offers the deployment advantages of not introducing any additional data overhead, as well as being entirely supported by current network protocols and hardware.
We do not consider defenses such as BuFLO or HTTPOS given their complexity, cost and the effectiveness of the alternatives we do consider~\cite{dyer,luo}.

The exponential defense slightly outperforms the linear defense, decreasing the accuracy of the Pan attack from 60\% to 22\% and the BoG attack from 89\% to 59\%.
Notice that the exponential defense is much more effective in our evaluation context than Dyer's context, which focuses on comparing website homepages loaded over an SSH tunnel with caching disabled and evaluation traffic collected on the same machine as training traffic.
The fragmentation defense is extremely effective against the LL and Wang - FLL attacks, reducing accuracy to below 1\% for each attack, but less effective against the Pan and BoG attacks.
The Pan and BoG attacks each perform TCP stream re-assembly, aggregating fragmented packets, while LL and Wang - FLL do not.
Since TCP stream re-assembly is expensive and requires complete access to victim traffic, the fragmentation defense may be a superior choice against many adversaries in practice.

Although the fragmentation, linear and exponential defenses offer the deployment advantages of functioning statelessly below the TCP layer, their effectiveness is limited.
The Burst defense offers greater protection, operating between the TCP layer and application layer to pad contiguous bursts of traffic up to pre-defined thresholds uniquely determined for each website.
The Burst defense allows for a natural tradeoff between performance and cost, as fewer thresholds will result in greater privacy but at the expense of increased padding.

Unlike the BoG attack which considers bursts as a tuple, padding by the Burst defense is independent in each direction.
We determine Burst thresholds as shown in Algorithm~\ref{alg:thresholds}, repeating the algorithm for each direction.
We pad traffic bursts as shown in Algorithm~\ref{alg:padding}.

\algrenewcommand\algorithmicrequire{\textbf{Precondition:}}

\begin{algorithm}
\caption{Threshold Calculation
\label{alg:thresholds}}
\begin{algorithmic}[1]
\Require{$bursts$ is a set containing the length of each burst in a given direction in defense training traffic}
\Require{$threshold$ specifies the maximum allowable cost of the Burst defense}
\State $thresholds \gets set()$
\State $bucket \gets set()$
\For{$b$ \textbf{in} \texttt{sorted}$(bursts)$}
   \State $inflation \gets \texttt{len}(bucket + b) * \texttt{max}(bucket + b) / \texttt{sum}(bucket + b)$
   \If {$inflation \geq threshold$}
   	\State $thresholds \gets thresholds + max(bucket)$
	\State $bucket \gets set() + b$
   \Else
   	\State $bucket \gets bucket + b$
   \EndIf
\EndFor
\State \Return $thresholds + \texttt{max}(bucket)$
\end{algorithmic}
\end{algorithm}
\begin{algorithm}
\caption{Burst Padding
\label{alg:padding}}
\begin{algorithmic}[1]
\Require{$burst$ specifies the size of a directed traffic burst}
\Require{$thresholds$ specifies the thresholds obtained in Algorithm~\ref{alg:thresholds}}
\For{$t$ \textbf{in} \texttt{sorted}$(thresholds)$}
   \If {$t \geq burst$}
       \State \Return $t$
   \EndIf
\EndFor
\State \Return $burst$
\end{algorithmic}
\end{algorithm}

We evaluate the Burst defense for $threshold$ values 1.10 and 1.40, with the resulting cost and performance shown in Figure~\ref{fig:def_performance}.
The Burst defense outperforms defenses which operate solely at the packet level by obscuring features aggregated over entire TCP streams.
Simultaneously, the Burst defense offers deployability advantages over techniques such as HTTPOS since the Burst defense is implemented between the TCP and application layers.
The cost of the Burst defense compares favorably to defenses such as HTTPOS, BuFLO and traffic morphing, reported to cost at least 37\%, 94\% and 50\% respectively~\cite{cai,wright}.
Having demonstrated the performance and favorable cost of the Burst defense, we plan to address further comparative evaluation in future work.

\section{Discussion and Conclusion}
\label{sec:conclusion}

This work examines the vulnerability of HTTPS to traffic analysis attacks, focusing on evaluation methodology, attack and defense.
Although we present novel contributions in each of these areas, many open problems remain.

Our examination of evaluation methodology focuses on caching and user-specific cookies, but does not explore factors such as browser differences, operating system differences, mobile/tablet devices or network location.
Each of these factors may contribute to traffic diversity in practice, likely degrading attack accuracy.
In the event that differences in browser, operating system or device type negatively influence attack results, we suggest that these differences may be handled by collecting separate training data for each client configuration.
We then suggest constructing a HMM that contains isolated site graphs for each client configuration.
During attack execution, the classifier will likely assign higher likelihoods to samples from the client configuration matching the actual client, and the HMM will likely focus prediction within a single isolated site graph.
By identifying the correct set of training data for use in prediction, the HMM may effectively minimize the challenge posed by multiple client configurations.
We leave refinement and evaluation of this approach as future work.

Additional future work remains in the area of attack development.
To date, all approaches have assumed that the victim browses the web in a single tab and that successive page loads can be easily delineated.
Future work should investigate actual user practice in these areas and impact on analysis results.
For example, while many users have multiple tabs open at the same time, it is unclear how much traffic a tab generates once a page is done loading.
Additionally, we do not know how easily traffic from separate page loadings may be delineated given a contiguous stream of user traffic.
Lastly, our work assumes that the victim actually adheres to the link structure of the website.
In practice, it may be possible to accommodate users who do not adhere to the link structure by introducing \textit{strong} and \textit{weak} transitions rather than a binary transition matrix, where strong transitions are assigned high likelihood and represent actual links on a website and weak transitions join all unlinked pages and are assigned low likelihood.
In this way the HMM will allow transitions outside of the site graph provided that the classifier issues a very confident prediction.

Defense development and evaluation also require further exploration.
Attack evaluation conditions and defense development are somewhat related since conditions which favor attack performance will simultaneously decrease defense effectiveness.
Defense evaluation under conditions which favor attack creates the appearance that defenses must be complex and expensive, effectively discouraging defense deployment.
To increase likelihood of deployment, future work must investigate necessary defense measures under increasingly realistic conditions since realistic conditions may substantially contribute to effective defense.

This work has involved substantial implementation, data collection and computation.
To facilitate future comparative work, our data collection infrastructure, traffic samples, attack and defense implementations and results are available upon request.

\bibliographystyle{IEEEtran}
\bibliography{biblio}

\begin{figure}[t]
        \begin{center}
                \label{fig:diversity}
                \includegraphics[width=\columnwidth]{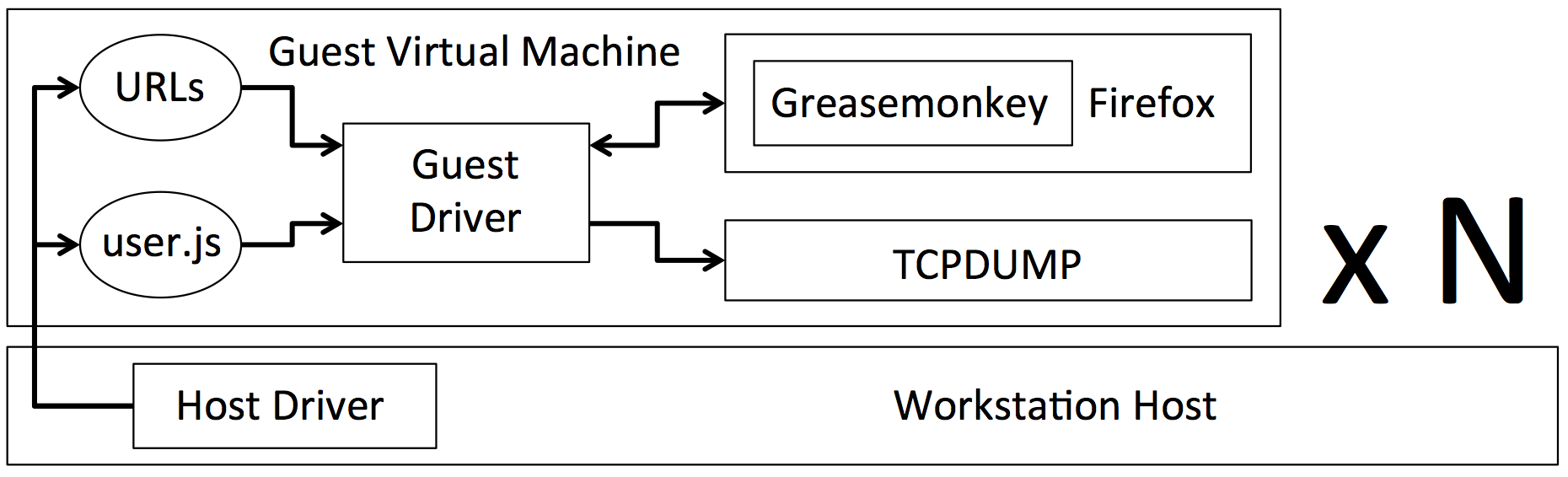}
        \end{center}
        \caption{Data collection infrastructure.}
        \label{fig:apparatus}
\end{figure}

\section*{Appendix A}
\label{sec:aparatus}

In this appendix we describe the software system used to collect traffic samples.
To allow parallel collection of data, we collected traffic inside a VirtualBox 4.2.16 VM running Linux 12.04 and Firefox 22.
We ran 4 VMs at a time on the same workstation with a quad core Xeon processor and 12GB of RAM.
We ran only 4 VMs to allow each VM sufficient processor, memory, disk and network resources to prevent any dropped packets or delays in website loading.
Since some collection modes require a fresh VM image for each browsing session, we disabled automatic updates in Ubuntu and Firefox as these would have consistently downloaded updates and contaminated traffic samples.

Figure~\ref{fig:apparatus} depicts the software components of the collection infrastructure.
The HostDriver managed the collection process, including booting VMs and assigning workloads.
The HostDriver shared a folder with each VM containing a \texttt{user.js} file used to configure Firefox caching behavior and list of URLs to visit.
To disable the Firefox cache, we set 15 caching related configuration options listed in Firefox source file \texttt{all.js}.
Each VM launched the GuestDriver script after boot, which then launched Firefox using the supplied \texttt{user.js} configuration file.
A Greasemonkey script (version 0.9.20) installed in Firefox then successively visited each of the listed URLs.
After each webpage had fully loaded the Greasemonkey script waited 3 seconds to allow for any JavaScript URL redirection.
Once any redirection had finished, the script waited an additional 4 seconds to ensure all traffic generated by the page was collected.
Greasemonkey then issued a blocking request to a server running locally on the VM which caused the server to stop and restart TCPDUMP, thereby creating separate PCAP files for each sample.
Note that we disabled Content Security Policy as several sites had policies which prevented our Greasemonkey script.

\section*{Appendix B}
\label{sec:canonicalization}
In this appendix we further describe the techniques we use to produce the canonicalization function.
Recall that the canonicalization function transforms the URL displayed in the browser address bar after a webpage loads into a label identifying the contents of the webpage.

The most basic heuristics we apply in the canonicalization function handle differences in the URL which rarely impact content.
We remove the \texttt{www} subdomain at the beginning of the full domain name, convert all URLs to be lower case and remove any trailing ``\texttt{/}'' at the end of the URL.
Beyond these, we assume that any two URLs which differ prior to the query string will correspond to webpages which are not the same.
This assumption is not inherent to the concept of a canonicalization function and could be removed by modifying the canonicalization function to also operate on the domain, port and path of a URL.

Having canonicalized the domain, port and path of the URL, we now identify any arguments appearing in the query string which appreciably impact page content.
We enumerate all arguments which appear in any URL at the website, and then for each argument enumerate all values associated with that argument and a list of all URLs in which each (argument, value) pair appears.
We then iterate through arguments to identify arguments that significantly impact page content.
For each argument, we randomly select 6 URL paths in which the argument appears and up to six distinct values of the argument for each URL path.
Note that the impact of an argument can normally be determined by viewing simply a pair of argument values for a single URL path; we consider additional samples as described below to provide a margin of safety.

The decision process for deciding whether a particular argument significantly influences content is as follows:
If all pages with the same base URL appear the same, then the argument does not influence content.
If pages with the same base URL appear different, and the argument being examined is the only difference in the URL, then the argument does influence content.
In the case that pages with the same base URL do appear different and multiple arguments are different, additional investigation is necessary.
If removal of the argument causes page content to change, then the argument influences page content.
Alternately, if substitution of alternate argument values causes page content to change, then the argument influences page content.
Once we have identified all arguments which do not impact page content, we canonicalize URLs by removing these arguments and their associated values.

This approach to canonicalization makes several assumptions.
The approach assumes that the impact of the argument is independent of the URL path.
Additionally, the approach assumes that the effect of each argument can be observed by manipulating that argument independent of any other arguments.
To provide limited validation of our assumptions, we perform a ``safety check'' for each website which randomly selects labels and compares URLs corresponding to the label to verify that page contents are comparable.

\end{document}